\documentclass[sigconf]{acmart}

\usepackage{multirow}
\usepackage{subfiles}
\usepackage{subfigure}
\usepackage{mathbbol}
\usepackage{enumitem}

\AtBeginDocument{%
  \providecommand\BibTeX{{%
    \normalfont B\kern-0.5em{\scshape i\kern-0.25em b}\kern-0.8em\TeX}}}

\makeatletter
\let\@authorsaddresses\@empty
\makeatother

\copyrightyear{2023}
\acmYear{2023}
\setcopyright{rightsretained}
\acmConference[CIKM '23]{Proceedings of the 32nd ACM International Conference on Information and Knowledge Management}{October 21--25, 2023}{Birmingham, United Kingdom}
\acmBooktitle{Proceedings of the 32nd ACM International Conference on Information and Knowledge Management (CIKM '23), October 21--25, 2023, Birmingham, United Kingdom}\acmDOI{10.1145/3583780.3614775}
\acmISBN{979-8-4007-0124-5/23/10}

\begin{document}
\newcommand{\preedit}[1]{\textcolor{olive}{#1}}
\newcommand{\holden}[1]{{\bf \color{magenta} [[$_{holden}$ ``#1'']]}}
\newcommand{\reedit}[1]{\textcolor{black}{#1}}

\title{Adaptive Multi-Modalities Fusion in Sequential Recommendation Systems}

\author{Hengchang Hu}
\email{hengchang.hu@u.nus.edu}
\authornote{Work done when the author is a research intern at Huawei Noah's Ark Lab, Singapore.}
\authornote{Corresponding author.}
\affiliation{%
  \institution{National University of Singapore}
  \country{Singapore}
}

\author{Wei Guo}
\email{guowei67@huawei.com}
\affiliation{%
  \institution{Huawei Noah’s Ark Lab}
  \country{Singapore}
}

\author{Yong Liu}
\email{liu.yong6@huawei.com}
\affiliation{%
  \institution{Huawei Noah’s Ark Lab}
  \country{Singapore}
}

\author{Min-Yen	Kan
}
\email{kanmy@comp.nus.edu.sg}
\affiliation{%
  \institution{National University of Singapore}
  \country{Singapore}
}

\renewcommand{\shortauthors}{Hengchang Hu,Wei Guo, Yong Liu, \& Min-Yen Kan}

\newcommand{\systemname}{MMSR}
\begin{abstract}
In sequential recommendation, multi-modal information (e.g., text or image) can provide a more comprehensive view of an item's profile. 
The optimal stage (early or late) to fuse modality features into item representations 
is still debated.
We propose a graph-based approach (named MMSR) to fuse modality features in an adaptive order, enabling each modality to prioritize either its inherent sequential nature or its interplay with other modalities.
MMSR represents each user's history as a graph, where the modality features of each item in a user's history sequence are denoted by cross-linked nodes.
The edges between homogeneous nodes represent intra-modality sequential relationships, and the ones between heterogeneous nodes represent inter-modality interdependence relationships.
During graph propagation, MMSR incorporates dual attention, differentiating homogeneous and heterogeneous neighbors.
To adaptively assign nodes with distinct fusion orders, MMSR allows each node's representation to be asynchronously updated through an update gate.
In scenarios where modalities exhibit stronger sequential relationships, the update gate prioritizes updates among homogeneous nodes. Conversely, when the interdependent relationships between modalities are more pronounced, the update gate prioritizes updates among heterogeneous nodes. Consequently, MMSR establishes a fusion order that spans a spectrum from early to late modality fusion.
In experiments across six datasets, MMSR consistently outperforms state-of-the-art models, and our graph propagation methods surpass other graph neural networks. Additionally, MMSR naturally manages missing modalities.
The code is available at: \url{https://github.com/HoldenHu/MMSR}. 
\end{abstract}



\maketitle

\section{Introduction}
\label{sec:intro}
\begin{figure}[t]
    \centering
    \setlength{\abovecaptionskip}{-0cm}
    \setlength{\belowcaptionskip}{-0.5cm}
    \includegraphics[width=0.48\textwidth]{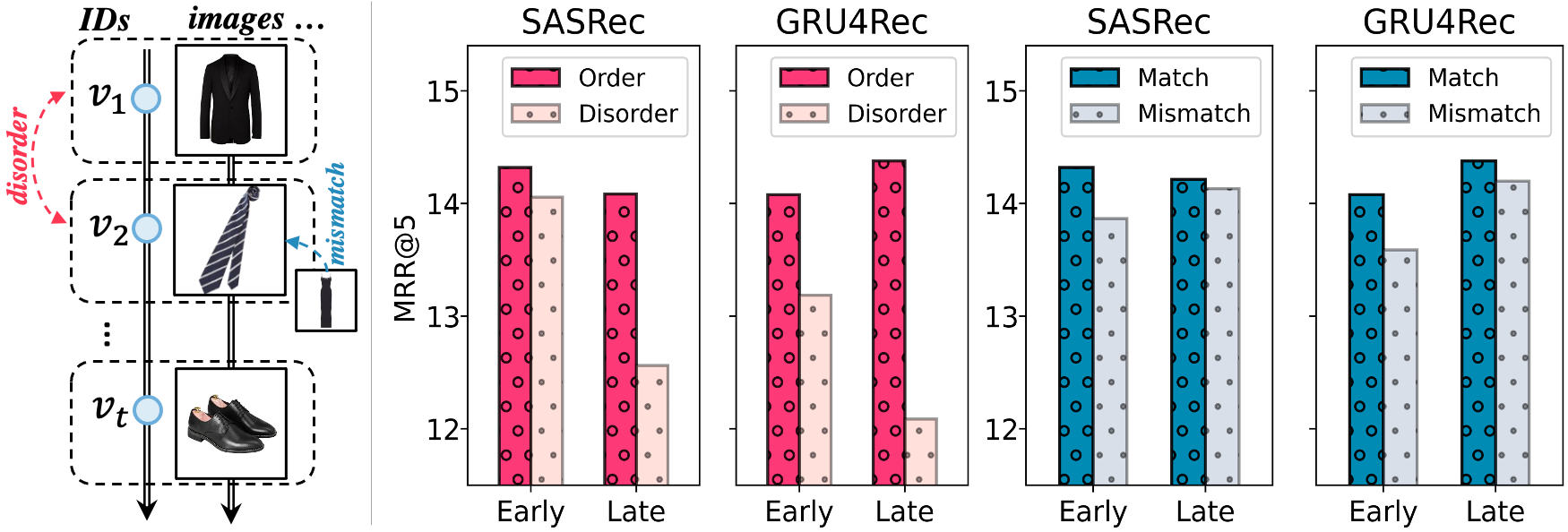}
  \caption{Case study on the Amazon-Fashion dataset. Here, \textit{Order}/\textit{Match} refers to the original modality sequence, while \textit{disordered} refers to a shuffled item order sequence, and \textit{mismatched} refers to a condition with displaced modalities. 
  }
    \label{fig:intro}
\end{figure}

Recommendation systems leverage user--item interactions to predict future user consumption. Collaborative approaches focus on determining similarity between users/items, while sequential approaches uncover sequential patterns among items. Modality information, such as images or text, has been extensively studied in collaborative recommendation \cite{wei2019mmgcn,tao2020mgat,chen2022breaking,zhou2023bootstrap}, but its potential in sequential recommendation (SR) remains largely unexplored.
In collaborative recommendation, modalities are represented as high-dimensional feature vectors, which are captured through pre-trained models like BERT \cite{devlin2018bert} for texts and ResNet \cite{he2016deep} for images. However, incorporating multiple modalities into SR poses two key challenges:
\textbf{(1)} Identifying sequential patterns within each modality, as they may exhibit distinct patterns;
\textbf{(2)} Capturing the complex interplay between modalities that can influence users' sequential behavior.
For example,  many consumers may purchase a \textit{suit} and then subsequently buy a \textit{tie} (Figure~1, left). Recognizing meaningful sequential image patterns between suits and ties allows for robust recommendations, independent of specific ID patterns.
Moreover, at the item-level, an item is not solely defined by a single modality. Considering different images of \textit{suits}, the interaction between these images and other modalities (e.g., textual descriptions) also plays a role in influencing a user's selection.

In Sequential Recommendation, existing approaches for merging different channels of features include early \cite{hidasi2016parallel,rashed2022context,lei2019tissa} and late fusion \cite{zhang2019feature}, which determine whether merging occurs before or after sequential modeling.
However, considering the above challenges, both have limitations --- early fusion is less sensitive to the interactions between intra-channel features, while late fusion is less sensitive to the interactions among different 
channels of features. 

We conduct a case study as evidence.  We utilized both fusion strategies on GRU4Rec \cite{hidasi2015session} and SASRec \cite{kang2018self} for \reedit{pre-experiments} on the Amazon dataset \cite{he2016ups}. To minimize interference, we had two settings: randomly shuffling the item-level sequence (\textit{disordered}) and maintaining the sequence while randomly displacing certain modalities (\textit{mismatched}).
We found late fusion models are more sensitive to the \textit{disordered} version (resulting in a significant performance drop). In contrast, early fusion is less sensitive to sequential patterns within each channel. Under \textit{mismatched} conditions, this reversed, with early fusion experiencing a larger performance drop. This indicates that late fusion is less sensitive to restricted modality matching.


These findings reveal that \textbf{fusion order is crucial}.
While holistic fusion methods like Trans2D \cite{singer2022sequential} suggest features can be fused without a strict order, they do not address the heterogeneity of feature channels or consider fusion order impact. Therefore, we propose a graph-based holistic fusion method for a flexible modality feature fusion.
As existing fusion methods 
target attribute features \cite{liu2021noninvasive,xie2022decoupled}, we introduce our Multi-Modality enriched Sequential Recommendation (MMSR) framework which focuses on modality feature fusion. 
Our MMSR framework comprises three stages: {\it representation}, where item features in each channel are represented as nodes; {\it fusion}, which aggregates features from different channels using graph techniques; and {\it prediction}, which generates the final representations.
To overcome the limitations of existing methods, we aim to tackle the aforementioned two challenges by: \textbf{(1)} Preserving modalities' temporal order during fusion, and \textbf{(2)} Enabling effective interactions between multiple modalities.

We represent each user's behavior history with a graph, where the modality features of items are nodes. We consider three feature channels: item identifier, visual, and textual modalities. Each graph maintains their temporal order as homogeneous relations while capturing cross-modal interactions as heterogeneous relations. Still, challenges persist in graph construction, aggregation, and updating.

\textit{Firstly}, in graph construction, treating each modality (such as images) as an individual node will overlook their semantic relatedness.
Moreover, given the three channels, the number of nodes in the graph processing will triple, significantly increasing the graph's sparsity.
\textit{Secondly}, graph nodes and relations are typed. During graph aggregation, simply viewing them as homogeneous nodes and relations results in oversimplification, resulting in poor representation and confusing fusion order (similarly, invasive feature fusion across channels also disrupts graph aggregation \cite{liu2021noninvasive}).
\textit{Thirdly}, na\"{\i}ve graph updating is synchronous for all nodes, unable to support fusion order.

To tackle these issues, we propose solutions.
\textit{Firstly}, to construct graphs, we adopt a similar approach \cite{yu2014factor} to create compositional embeddings that represent nodes as compositions of smaller groups.
Specifically, we cluster modality features and select the identifiers of the cluster centers as modality codes, which are then treated as new nodes in the graph.
This approach offers a two-fold advantage: reducing 
overfitting by having fewer modality nodes during training, and establishing 
links between items by grouping highly similar modalities under the same node.
\textit{Secondly}, for graph aggregation, we employ a dual attention function that distinguishes between homogeneous and heterogeneous nodes' correlations.  This utilizes content-based attention and key-value attention for measurement, respectively. 
\reedit{
Expanding on this, we propose a non-invasive propagation method that allows homogeneous and heterogeneous neighbors to influence --- but not invasively disrupt --- each other.}
\textit{Thirdly}, for graph updating, in MMSR, each node adaptively chooses the order of fusion through an update gate. This means each node can decide whether to fuse heterogeneous information first followed by homogeneous information, or vice versa. 

We experiment across six diverse scenarios, incorporating both image and text modalities as feature sets. 
\reedit{
In ablation experiments, we found that the optimal order for modality fusion --- whether early or late --- varies
per dataset.
Our proposed method, which adaptively determines the fusion order for each node, 
strikes balance,
consistently enhancing the efficacy of fusion.}
Our MMSR outperforms the 
state-of-the-art baselines by 8.6\% in terms of HR@5 on average, while also 
exhibiting strong robustness to missing modalities in real-world scenarios. 
We show that this is because MMSR enables items to search for matching visual or linguistic features, even in the absence of certain text or image nodes, rather than simply replacing missing modalities with default values.
\reedit{
MMSR can be scaled beyond two modalities, and thus is 
practical 
for diverse real-world multi-modal scenarios.
}

We summarise our contributions as follows:
\textit{(i)} We spotlight challenges in modality fusion for sequential recommendation, and propose a versatile solution --- our MMSR framework. It accommodates both early and late fusion across modalities.
\textit{(ii)} We offer a graph-centric holistic fusion method as the engine in MMSR, enabling the adaptive selection of fusion order for each feature node.
\textit{(iii)} We conduct comprehensive experiments on six datasets, which show significant gains in both accuracy and robustness.

\section{Related Work}

\subsection{Multi-modal Recommendation}

Multimodal recommendation systems leverage features from various modalities, including textual content \cite{rao2013entity, hu2022modeling, gong2016hashtag, zheng2017joint, wu2022mm, malkiel2020recobert} and images \cite{niu2018neural, deldjoo2021study, deldjoo2018audio, luo2008personalized, van2013deep, oramas2016sound}, to enhance item representations. 
Image feature extraction encompasses signal detectors for color \cite{yu2018aesthetic} and texture \cite{chi2016ubishop}; local feature extractors for detecting objects in coherent regions \cite{chi2016ubishop, gu2016iglasses}; and latent feature encoders using pre-trained CNN models \cite{li2020hierarchical, geng2015learning}. Textual features range from concept-level features (obtained via tools like NER \cite{rao2013entity} and TextRank \cite{hu2022modeling}), to semantic features from encoders like CNN \cite{gong2016hashtag, zheng2017joint} or  BERT \cite{wu2022mm, malkiel2020recobert}.

To obtain the final hidden representation, 
the fusion \cite{zhou2023comprehensive} of modality features can occur either before \cite{he2016vbpr,liu2019user} or after \cite{wei2019mmgcn,tao2020mgat} being sent into the feature interaction module.
Approaches such as MGAT \cite{tao2020mgat} directly sum the features to disentangle personal interests by modality and aggregate them into the final item representation. MMGCN \cite{wei2019mmgcn} merges modality-specific graphs through concatenation, but may not fully capture intermodal relations. In contrast, EgoGCN \cite{chen2022breaking} introduces Ego fusion, extending information propagation beyond the unimodal graph to capture relationships between modalities. It aggregates informative intermodal messages from neighboring nodes, generating final representations by combining multimodal and ID embedding propagation results via concatenation.
Despite these advancements, current multi-modal recommendation research predominantly targets collaborative tasks, still leaving the use of multi-modality in sequential recommendation largely unexplored.

\subsection{Feature Fusion in Sequential Recommenders}

\reedit{
Sequential recommenders (such as GRU- \cite{hidasi2015session}, Transformer- \cite{kang2018self}, or BERT-based \cite{sun2019bert4rec} models) capture user interests using item ID sequences. To incorporate additional item features (primarily attribute features), fusion methods are used to integrate them into the overall item representation. These fusion methods can be categorized as late, early, or holistic fusion,  depending on {\bf when} feature representations are merged.
}

In \textit{late fusion}, sequential relationships within each feature channel are modeled before merging them in a final stage. 
For example, FDSA \cite{zhang2019feature} separately encodes item and side features using self-attention before fusion.  Conversely, early fusion integrates feature representations prior to exploring sequential interactions. \textit{Early fusion} can be invasive or non-invasive. Invasive methods irreversibly merge item IDs with side features through techniques like concatenation \cite{rashed2022context,tang2018personalized}, addition \cite{hidasi2016parallel,sun2019bert4rec}, or gating \cite{lei2019tissa}.
As an example, DETAIN \cite{lin2022sequential} uses a 2D approach to handle sequential items' features, merging feature channels with vertical attention and items with horizontal attention.
However, these methods alter the original representations and have documented drawbacks in terms of compound embedding space \cite{liu2021noninvasive}. Non-invasive approaches do not directly mix item representation with features. For example, NOVA \cite{liu2021noninvasive} fuses features while maintaining consistency in item representation. DIF-SR \cite{xie2022decoupled} introduces an attribute-based attention function for fusing items. 
In contrast, \textit{Holistic fusion} posits that modality fusion and sequential modeling can proceed without rigid ordering. 
Trans2D \cite{singer2022sequential} employs 4D attention matrices to gauge item attribute correlations but overlooks the ordering of heterogeneous and homogeneous relations.
Our work introduces an adaptive method that determines relation application order per node during propagation, providing a more versatile solution.

\section{Preliminaries}

In our problem, the core task is 
sequential recommendation: Given a user $u$'s historical interaction data $\mathbf{H}_u$, the aim is to find a function $f: \mathbf{H}_u \rightarrow v$ that predicts the next item $v$ that the user is most likely to consume.
In a typical sequential recommendation task, the historical interaction data includes only item ID information; i.e., $\mathbf{H}_u = \{v_1, v_2, \dots, v_m\}$. Based on this foundation, modality-enhanced sequential recommendation considers the modality of items in the sequence as well, represented by $\mathbf{H}_u = \{\mathbb{x}_{1}, \mathbb{x}_{2}, \dots, \mathbb{x}_{m}\}$, where each $\mathbb{x}$ is the combination of different 
feature channels of the item (including item identifier and item modalities). In this work, we only consider image and text modalities (although extensible to other modalities), and one instance is represented as $\mathbb{x}_i: \{v_i, a_i, b_i\}$. Here, $a_i$ and $b_i$ indicate the image and text of item $v_i$, respectively.
To simplify our discussion, we will refer item ID, image feature, and text feature as three feature channels of modalities; i.e., $v \in \mathcal{V}$, $a \in \mathcal{A}$, and $b \in \mathcal{B}$.

\subsection{Base Model}
We now discuss the base sequential recommendation model, which we characterize as a 3-tuple of (an Embedding, Representation learning , Prediction). 

\paragraph{Initial embedding}
The item ID features are represented as integer index values and can be converted into low-dimensional, dense real-value vectors by performing table lookups from an embedding table.
For modality embeddings, the commonly-used approach is to directly utilize its extracted features and represent them as a feature vector, through a third-party model \cite{he2016deep,raffel2020exploring}.
In order to obtain a comprehensive embedding tensor $\mathbf{E} \in \mathcal{R}^{3 \times m \times d}$ of the input features of user history, the feature channels are organized in columns and the sequences are organized in rows.
\begin{equation}
\mathbf{E} =
    {\left[ 
    \begin{array}{cccc}
        \mathbf{e}_{v_1}, & \mathbf{e}_{v_2}, & \cdots, & \mathbf{e}_{v_m} \\
        \mathbf{e}_{a_1}, & \mathbf{e}_{a_2}, & \cdots, & \mathbf{e}_{a_m} \\
        \mathbf{e}_{b_1}, & \mathbf{e}_{b_2}, & \cdots, & \mathbf{e}_{b_m}
    \end{array}
    \right ]}
\end{equation}

\paragraph{Representation learning}
Numerous existing works have concentrated on designing network architectures for the purpose of modeling feature interactions, outputting the user representation $\mathbf{P}$. This can be expressed as:
\begin{equation}
    \mathbf{P} = f(\mathbf{E})
\end{equation}
For early fusion, the vertical feature channels are fused first, followed by the fusion of the horizontal sequence relationships.
For simplicity, we use a linear combination for fusion through channels.
\begin{align}
    \mathbf{E}_{i,:} = \sigma(\mathbf{W} (cat[\mathbf{E}_{i,1}; \mathbf{E}_{i,2}; \mathbf{E}_{i,3}]))  \\
    \mathbf{P} = \mathcal{M} ([\mathbf{E}_{1,:}, \mathbf{E}_{2,:}, ..., \mathbf{E}_{m,:}])
\end{align}
For late fusion, the order is reversed and can be formulated as:
\begin{align}
    \mathbf{E}_{:,j} = \mathcal{M} ([\mathbf{E}_{1,j}, \mathbf{E}_{2,j}, ..., \mathbf{E}_{m,j}])   \\
    \mathbf{P} = \sigma(\mathbf{W} (cat[\mathbf{E}_{:,1}; \mathbf{E}_{:,2}; \mathbf{E}_{:,3}]))
\end{align}
\noindent where $cat[;]$ is the concatenation operation, $\mathbf{W}$ is the linear weight parameter, $\sigma$ is the activation function, and $\mathcal{M}$ is the models for sequence modeling.
In contrast, for holistic fusion, the $f$ will process $\mathbf{E}$ as a whole. Trans2D \cite{singer2022sequential} direct applies 2D-attention on $\mathbf{E}$, and our method considers $\mathbf{E}$ as node representations in a graph structure.

\paragraph{Prediction}
By scoring candidates items $\mathbf{e}_{v}$ against the learned user representation $\mathbf{P}$ using a dot product, we generate the predicted probability scores:

\begin{equation}
    \hat{y} = <\mathbf{P}, \mathbf{e}_{v}^\top>
\end{equation}
During training, the model measures and minimizes the differences between the ground-truth $y$ and the prediction $\hat{y}$ through cross-entropy loss \cite{kang2018self}.

\section{Approach}

As stated earlier, the fusion order during the representation learning stage is crucial. Current methods fail to balance the extremes of the two orders. To address this, we propose the MMSR framework, which extends the base model and incorporates a graph-based fusion neural network in the representation learning stage to fuse features. 
After constructing Multi-modal Sequence Graphs for each user, we utilize a dual attention mechanism to independently aggregate heterogeneous and homogeneous node information, enabling an adaptive merging order that facilitates simultaneous consideration of both sequential and cross-modal aspects.


\begin{figure*}[t]
    \centering
    \setlength{\abovecaptionskip}{-0cm}
    \setlength{\belowcaptionskip}{-0.3cm}
    \includegraphics[width=0.97\textwidth]{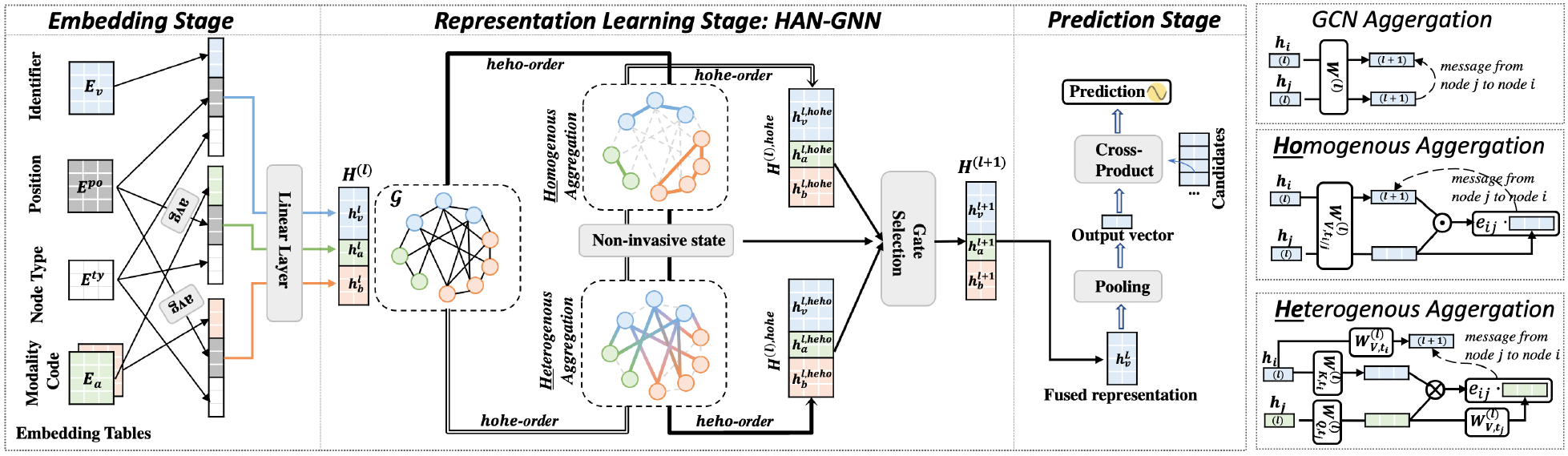}
  \caption{Overall framework of MMSR (left), and the applied aggregation modules (right). Distinct node types are represented by different colors.}
    \label{fig:framework}
\end{figure*}

\subsection{Multimodal Sequence Graph Construction}

\begin{figure}[t]
    \centering
    \setlength{\abovecaptionskip}{-0cm}
    \setlength{\belowcaptionskip}{-0.4cm}
    \includegraphics[width=0.45\textwidth]{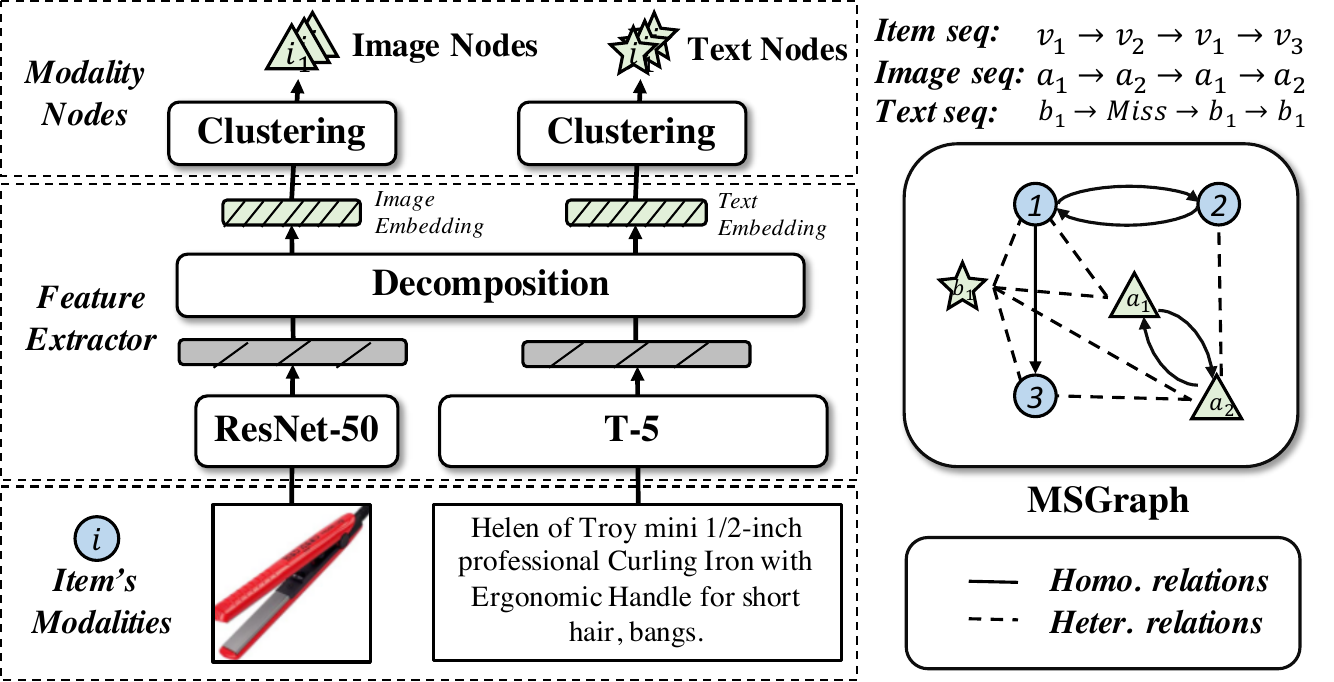}
  \caption{Modality-enriched graph construction. }
    \label{fig:cons}
\end{figure}

For each user $u$, we represent his/her history as a graph --- a Modality-enriched Sequence Graph (MSGraph), $\mathcal{G}_u = (\mathcal{N}_u, \mathcal{R}, \mathcal{E}_u)$. 
Note that each user's graph $\mathcal{N}_u$ and $\mathcal{E}_u$ can differ. For simplicity, we'll refer to a single user's graph, and just represent them as $\mathcal{N}$ and $\mathcal{E}$, in the discussion that follows.
Figure~\ref{fig:cons} depicts the construction pipeline. The right side illustrates node construction from modalities, while the left details the edge construction within the MSGraph.

\vspace{0.8mm}
\textit{Nodes and their initialization.}
Each MSGraph should consist of $m \times 3$ nodes (where $m$ is the sequence length), forming the node set $\mathcal{N}$. $\mathcal{N}$ encompasses the three types of nodes, representing three distinct features of channels: $\{{v_1},...,{v_m}\}$, $\{{a_1},...,{a_m}\}$, and $\{{b_1},...,{b_m}\}$. Their representations are associated with the first row (item ID feature), second row (image feature), and third row (text feature) of matrix representation tensor $\mathbf{E}$, respectively.

During node representation initialization, $\mathbf{e}_v$ is randomly initialized. For $\mathbf{e}_a$ and $\mathbf{e}_b$, we extract semantic features from the corresponding modality. Our method is not limited to image and text modalities, and for better extension ability, we use separate models instead of large visio-linguistic models for feature extraction. Visual features $\mathbf{e}_a$ are obtained from a ResNet-50 \cite{he2016deep} model pretrained on ImageNet \cite{deng2009imagenet}, while textual features $\mathbf{e}_b$ are extracted using a pretrained T-5 model \cite{raffel2020exploring}. This scheme can be represented as "modality $a$ $\Rightarrow$ representation $\mathbf{e}_a$" (the same applies to $b$).

\vspace{0.8mm}
\textit{Node transformation and compositions.}
According to \citet{hou2023learning}, closely binding text encodings with item representations can be detrimental. Thus, instead of using each modality as an individual node, we introduce \textit{``modality codes''} \cite{hou2023learning,rajput2018recommender} as alternative nodes. These nodes correspond to discrete indices obtained by mapping the original modality features. This approach helps alleviate the tight binding between item modality and item representations. The node representations utilize these indices to look up the code embedding table, resulting in a scheme denoted as ``modality $a$ $\Rightarrow$ code $ID_a$ $\Rightarrow$ representation $\mathbf{e}_a \simeq \mathbf{e}_{ID_a}$''.
To achieve this, we use a linear autoencoder \cite{baldi1989neural} to condense image/text feature vectors. We then use a K-means \cite{lloyd1982least} to cluster the modality feature vectors by modality type. 
The indices of cluster centers are used as modality codes $ID_a$. Initialized representations $\mathbf{e}_{ID_a}$ are derived from these cluster center representations.

We go beyond treating each item modality as an independent index, employing a composition technique. It enables mapping of multiple modalities to a single code, and a single modality to multiple codes. For example, both $a_1$ and $a_2$ can share a common modality node in the graph, and $a_1$ can correspond to multiple codes represented by $ID_{a_1}$, as a set of codes. 
By doing so, we significantly enhance the connectivity of features within each MSGraph.
To achieve this, we cluster each channel of modality into $c$ clusters and select the top $k$ nearest centers as the corresponding code set for each individual modality. The selection process is based on cosine similarity between the modality feature vectors and cluster center vectors. Here, $k$ represents a hyperparameter that determines the number of codes each modality is connected to.
For brevity, we will refer to the \textit{modality codes} as \textit{modalities}.

\vspace{0.8mm}
\textit{Edges and Relation Types.}
In the MSGraphs, we specify the edges as relations $\mathcal{E}$ between nodes, including \textit{homogeneous relations} $\mathcal{E}_{homo}$ and \textit{heterogeneous relations} $\mathcal{E}_{hete}$.
Both can be formulated as $\mathcal{E}: (n_s,r,n_o)$, indicating the relation $r$ between subject node $n_s$ and object node $n_o$ (where both $n \in \mathcal{N}$).
In $\mathcal{E}_{homo}$, $n_s$ and $n_o$ should be in the same type, such as encompassing items $(v,r,v)$ or modalities $(a,r,a)$. 
And the $r \in \mathcal{R}$ encompasses 3 types of sequential relations (intra-item or intra-modality): transition-in, transition-out, and bi-directional transitions.
The term ``\textit{transition}'' refers to the direct adjacent relationship in a sequence. For instance, if Item~A is selected immediately before Item~B, A to B is a transition-out relation, while B to A is a transition-in.
For modalities, we also establish direct connections between adjacent nodes in the sequence order. In case there is a back-and-forth relationship between the two modalities, we label it as a bi-directional relation.
In $\mathcal{E}_{hete}$, $n_s$ and $n_o$ belong to different node types, such as $(v,r,a)$ or $(a,r,b)$. There just exists one type of relation $r$, which signifies the correspondence matching between different feature channels of the same item. Additionally, in both types of relations, we introduce self-loop relations for each node to preserve its original information.

\subsection{Node Representation}


In MSGraph, each node is assigned an independent representation.
However, graphs pose a challenge when modeling sequential tasks as they undermine the inherent sequential nature \cite{chen2020handling}.
This issue is evident when graphs fail to reconstruct sequences due to repeated nodes, particularly as modality codes intensify this repetition. 
Additionally, the impact of different node types on user preferences within a sequence may vary. For example, images may have a more pronounced short-term influence on user preferences than text.

We propose a solution by 
integrating positional embeddings and node type embeddings into the original initialized representation $\mathbf{e}_n$ for each node.
These embeddings
map integer indices to low-dimensional dense vectors using separate embedding tables. 
Specifically, for position embedding of node $n$, its node type is embedded, yielding vector $\mathbf{e}^{ty}_n$.
Furthermore, the node's positions within the sequence are captured by a set of position indices, as modality nodes would take multiple positions. Each position index corresponds to an individual embedding, and the position embedding $\mathbf{e}^{po}_n$ is obtained by averaging these embeddings. This average vector indicates the position bias of the node towards the beginning or end of the sequence.
Finally, the node representation is combined as $\tilde{\mathbf{e}}_n = W [\mathbf{e}_n; \mathbf{e}^{ty}_n; \mathbf{e}^{po}_n]$, where $W$ is the weight parameter used for merging the concatenated embeddings.

\subsection{Representation Propagation Layers}
Given user graph $\mathcal{G}_{u}$, the next step involves aggregating the neighbor information for each node. This process can also be interpreted as \textbf{modal fusion}, where the sequential order and interdependencies between modalities are simultaneously taken into account.

\subsubsection{Synchronous Graph Neural Networks}
The most intuitive idea is to use graph neural networks to \textit{synchronously} fuse the node information together \cite{wei2019mmgcn,tao2020mgat,he2020lightgcn,ran2022pm}.
Here, \textit{synchronously} refers to all nodes being updated simultaneously from the previous layer to the next layer, without any specific order. 
Here, we denote the central node as $n_i$ and its corresponding neighbor set in the graph as $N_i$. The aggregator updates the representation of each node iteratively from the previous layer $h^{(l)}_i$ to the next layer $h^{(l+1)}_i$. 
Here, $h^{(0)}_i$ is initialized by $\tilde{\mathbf{e}}_i$.
We give examples of the following state-of-the-art graph aggregators as potential candidates to facilitate synchronous information propagation. 

\begin{itemize}[leftmargin=*]
    \item \textit{GCN Aggregator} \cite{kipf2016semi} takes into account the neighborhood information of a central node and aggregates it using a convolution operation. 
    Its formulation is represented as follows:
        \begin{equation}
            h_i^{(l+1)}=\sigma\left(\sum\nolimits_{j \in N_i} d(i,j) W^{(l)} h_j^{(l)} \right)
        \end{equation}
    \noindent where $\sigma$ and $W^{(l)}$ are the activation function and the transformation matrix of layer $l$.
    $d(i,j) = 1 / \sqrt{|N_i| |N_j|}$ is the normalization factor. We give an illustration in Figure~\ref{fig:framework} (upper right).
        
    \item \textit{GAT Aggregator} \cite{velickovic2017graph} further considers that each neighbor has a different impact on the central node, incorporating the attention mechanism to assign varying weights to neighbors: 
    \begin{equation}
        h_i^{(l+1)} = \sum\nolimits_{j \in N_i} \alpha_{ij}^{(l)} h_j^{(l)}
    \end{equation}
    \noindent \noindent where $\alpha_{ij}^{(l)}$ represents the attention score between node $i$ and node $j$. It is calculated by applying softmax to the dot product of a learnable weight vector $a$ and the concatenated representations of nodes $i$ and $j$ after linear transformations $W^{(l)}$:
    \begin{equation}
        e_{ij}^{(l)} = a^T [W^{(l)} h_i^{(l)} ; W^{(l)} h_j^{(l)}]
    \end{equation}
    \begin{equation}
        \alpha_{ij}^{(l)} = sft(e_{ij}^{(l)}|{N}_i) =\frac{\exp(\text{LeakyReLU}(e_{ij}^{(l)}))}{\sum_{k \in N_i} \exp(\text{LeakyReLU}(e_{ik}^{(l)}))}
    \end{equation}
    \noindent Here, $a$ is the parameter for calculating the attention score $e$. 
    For simplicity, we denote the softmax operation as $sft(e_{ij}^{(l)}|{N}_i)$.
\end{itemize}

\noindent As aggregators are very important for our method's performance, 
acting as the modality fusion module, we study the effectiveness of the above aggregators as well as other aggregators in the experiment section (\S~\ref{ss:gas}).

\subsubsection{Our Graph Neural Network}

There are some drawbacks to using the above graph neural networks:
\textit{Firstly}, concerning the 3 types of nodes and 5 types of relations, the heterogeneity of both should be taken into account.
\textit{Secondly}, as stated earlier, the order of fusion matters, thus synchronous updating is not optimal. Some prior modal information may be more beneficial to the corresponding item representation, thus should be merged first.
\textit{Thirdly}, the representations of the item, image, and text nodes are not in the same space, and thus are inappropriate to fuse them directly. 
The inclusion of different modalities can interfere with each other's representations, resulting in invasive fusion problems \cite{liu2021noninvasive}. This issue also persists during aggregation.
Based on these considerations, we propose an \textbf{H}eterogeneity-aware, \textbf{A}synchronous, and \textbf{N}on-invasive graph neural network (or HAN-GNN for short). 

\vspace{0.8mm}
\noindent \textbf{Heterogeneity-aware.}
To aggregate homogeneous and heterogeneous neighbor nodes, we employ a divide-and-conquer strategy.
For homogeneous nodes $n_j \in N_i$, they represent neighbors that are connected to the central node $n_i$ through edges $(n_i,r,n_j) \in \mathcal{E}_{homo}$, with their identical types $t_j = t_i$ (where $t$ belongs to the indices of three channels).
For heterogeneous nodes $n_j \in N_i$ with $t_j \neq t_i$, they represent neighbors that differ in type from the central node. These nodes are connected through edges $(n_i,r,n_j) \in \mathcal{E}_{hete}$.
Unlike GCN/GAT that transform the original hidden vector $h$ into a value vector using $W$ for layer-wise aggregation, our method establishes $W_Q$, $W_K$, and $W_V$ to transform the original vector into query, key, and value vectors, respectively. Considering three distinct node type $t$, three sets of parameters ($W_{Q,t}$, $W_{K,t}$, $W_{V,t}$) are designated.


For attention regarding \textit{homogeneous nodes}, their shared space allows direct comparison. We employ content-based attention for this. Formally, for $(n_i, r, n_j) \in \mathcal{E}_{homo}$, their attention scores are:
\begin{equation}
    e_{ij}^{(l),ho} = a_r (W_{V,t_i}^{(l)} h_i^{(l)} \odot W_{V,t_j}^{(l)} h_j^{(l)})  
\end{equation}

\noindent 
where $\odot$ is the element-wise production. $t_i$ and $t_j$ indicate the types of $n_i$ and $n_j$ respectively, which are 
identical in this case. It results in $W_{V,t_i} = W_{V,t_j}$.
Concerning types of sequential relations in $\mathcal{E}_{homo}$, for each relation $r$ between $n_i$ and $n_j$, we define an individual parameter $a_r$ for each relation type.

For the attention calculation with respect to \textit{heterogeneous nodes}, as they are located in different spaces, so we employ type-specific transformation matrices ($t_j \neq t_i$) to bring them into a common space for comparison. More specifically, we utilize key--value attention to evaluate the correlations between nodes after their individual transformations.
For $(n_i, r, n_j) \in \mathcal{E}_{hete}$, the attention scores are:
\begin{equation}
    e_{ij}^{(l),he} = (W_{Q,t_j}^{(l)} h_j^{(l)}) (W_{K,t_i}^{(l)} h_i^{(l)})^{\top}
\end{equation}

Building on the previously mentioned attention scores, we can independently gather the updated representations that each 
individually aggregate the two distinct sources of information. 
These can be succinctly expressed as follows:
\begin{equation}
    h_i^{(l+1), *} = \sum\nolimits_{j \in N_i} 
    sft(e_{ij}^{(l),*} | N_i) 
    (W_{V,t_j}^{(l)} h_j^{(l)})
\label{eq:aggregation}
\end{equation}
The symbol $*$ denotes either $ho$ or $he$, indicating the individual aggregation of homogeneous or heterogeneous information, respectively. This is represented as \textit{homogeneous aggregation} and \textit{heterogeneous aggregation} in Figure~\ref{fig:framework}~(right).

Providing information from both sources of neighbors, a direct approach would be to concatenate them and concurrently update this combined representation in the next layer:
\begin{equation}
    h_i^{(l+1)} = Linear([h_i^{(l+1), ho}; h_i^{(l+1), he}]),
\label{eq:asychronous}
\end{equation}
\noindent where $Linear$ indicates a linear layer. We refer to this approach as a ``\textit{synchronous} updating'' version of our implementation, to be evaluated in our ablation experiments (\S~\ref{ss:as}).

\vspace{0.8mm}
\noindent \textbf{Asynchronous updating.}
Synchronous updating overlooks the effect of the fusion order. Therefore, we propose an asynchronous updating strategy with two defined updating orders. 
In every layer $l$, one could either first aggregate homogeneous information for node updates and then use these updated representations for heterogeneous information aggregation, or vice versa. We term these two distinct updating orders as ``homogeneous-first, heterogeneous-second'' (or $hohe$) and ``heterogeneous-first, homogeneous-second'' ($heho$).
Taking $hohe$ as an example, this two-phase paradigm can be denoted as $h_i^{(l)} \rightarrow h_i^{(l),ho} \rightarrow h_i^{(l),hohe}$. 
The first phase $h_i^{(l)} \rightarrow h_i^{(l),ho}$ uses \textit{homogeneous aggregation} depicted in Equation~\ref{eq:aggregation}, but holding the layer number $(l)$ unchanged. The second phase $h_i^{(l),ho} \rightarrow h_i^{(l),hohe}$ follows the \textit{heterogeneous aggregation} in the Equation~\ref{eq:aggregation} but substitutes the input $h_i^{(l)}$ with the output $h_i^{(l),ho}$ from the first phase. 
Similarly, the $heho$ updating order follows: $h_i^{(l)} \rightarrow h_i^{(l),he} \rightarrow h_i^{(l),heho}$. 

As depicted in Figure~\ref{fig:framework} (left), in HAN-GNN, each node has two potential aggregation routes from its representation $h_i^{(l)}$ to $h_i^{(l+1)}$ via either the \textit{`heho'} or \textit{`hohe'} path. We introduce an update gate to adaptively select the optimal path for each node using the following gate selection mechanism:
\begin{equation}
    \beta = MLP([h_i^{(l+1), hohe}; h_i^{(l+1), heho}]),
\end{equation}
\begin{equation}
    h_i^{(l+1)} = \beta_0 \times h_i^{(l+1), hohe} + \beta_1 \times h_i^{(l+1), heho},
\end{equation}
\noindent where $MLP$ is a multi-layer perceptron, and $\beta \in \mathbb{R}^{2}$ contains the scores for gate selection.

\vspace{0.8mm}
\noindent \textbf{Non-invasive fusion.}
Drawing inspiration from NOVA \cite{liu2021noninvasive}, we employed a non-invasive technique to limit interference among different node types during feature updates.
For example, although the image features are fused with the item node in Phase~1, they do not actually update it but only use the updated representation for calculating the attention scores in Phase~2. 

Let us take the example of $h_i^{(l)} \rightarrow h_i^{(l),he} \rightarrow h_i^{(l),heho}$ to make this concept concrete. 
In Phase~1, the graph aggregates neighbors' information from the transformed value vector of $h_v^{(l)}$; while in Phase~2, the aggregation uses the transformed value vector of $h_i^{(l),he}$. This implies that the value vector used in the second phase has already undergone a substantial update.
Considering the degradation of representations caused by excessive fusion when modeling sequences with heterogeneous information, we introduce a non-invasive approach during graph updating. Specifically, in Phase~2, though we calculate the attention based on the intermediate state $h_i^{(l),he}$, we continue to use the value vector of $h_j^{(l)}$ (instead of $h_i^{(l),he}$) for aggregation. 
We posit that non-invasive technique also applies to the converse order for updating (i.e., $hohe$). Taking into consideration that each item is an independent entity, the question remains whether to permit the invasive integration of homogeneous contextual information into the fusion of heterogeneous information. We examine this question in our ablation experiments (\S~\ref{ss:as}).

\subsection{User Interest Representation and Prediction}
Following $L$ layers of aggregation, we obtain the final $h_v^{(L)}$ and set the collection of hidden states of item nodes $\{h_v^{(L)}, v \in \mathcal{N}_v\}$ forms the output $\mathbf{Z} \in \mathbb{R}^{|\mathcal{N}_v| \times d}$,
where $\mathcal{N}_v$ indicate item ID node set.
The resulting $\mathbf{Z}$ can be considered a representation that has undergone modal fusion. 
Hence, the key lies in the mapping function $\mathcal{P}: \mathbb{R}^{|\mathcal{N}_v| \times d} \rightarrow \mathbb{R}^{d}$ of outputting user representation $\mathbf{P} = \mathcal{P}(\mathbf{Z})$, facilitating the next-item prediction. From our observation, using graphs presents a challenge as it tends to diminish the impact of individual items, making it challenging to differentiate 
similar sequences. For instance, the graph model may produce similar representations for sequences such as $(v_1,v_2,v_3)$ and $(v_1,v_2,v_3,v_4)$. To address this issue,  instead of employing average pooling, 
we adopt 
\textit{last pooling}, where we select the last item from the sequence as the pooled representation. Specifically, we denote it as $\mathbf{P} = \mathbf{Z}_{|\mathbf{H}_u|}$.

\subsection{Model Comparison \& Complexity Analysis}

When sequential relationships between modalities are strong, the selection gate prioritizes updates among homogeneous nodes first. Conversely, when interdependent relationships are strong, the gate prioritizes updates among heterogeneous nodes. 
Thus, our framework can set a fusion order that spans from early to late modality fusion as a spectrum of possibilities, with $hohe$ representing late fusion and $heho$ early fusion.

For \textit{complexity comparison}, the fused representation from HAN-GNN (in the representation learning stage) can be used directly for online inference, matching the base model's time complexity. 
The main time cost for model training comes from layer-wise graph networks. Compared to GCN's complexity of $O(L|\mathcal{U}||\mathcal{E}_u|)$ and Graphormer's \cite{ying2021transformers} $O(L|\mathcal{U}||\mathcal{N}_u|^{2})$,  HAN-GNN takes $O(2L|\mathcal{U}||\mathcal{E}_u|)$ as there are two phases in each layer propagation.
Here, $|\mathcal{U}|$ indicates the number of users, and $|\mathcal{E}_u|$ and $|\mathcal{N}_u|$ indicate the average number of edges and nodes in each user graph, respectively.
While our approach is more complex than simpler networks like GCN, it offers lower complexity compared to yet more complex networks, like Graphormer, while still delivering superior performance. In the user graph, each node connects to its preceding and following nodes in the sequence, and at least 2 other modality nodes. With 4 edges per node, $|\mathcal{E}_u| = O(4|\mathcal{N}_u|) = O(\frac{4}{3}|\mathcal{H}_u|)$, where $\mathcal{H}_u$ signifies the user interaction sequence length.
Therefore, our method is more efficient than Graphormer when the user history exceeds $2 \times \frac{4}{3} = 2.66$. In typical cases where the average user history length varies between $7$ and $9$, our method is considerably more efficient.

\section{Experiment}

\textit{Datasets.}
In line with previous studies \cite{he2016vbpr,zhang2021mining}, we utilized the Amazon review dataset \cite{he2016ups} for evaluation. This dataset provides both product descriptions and images, with varying sizes across product categories. To showcase our approach's versatility, we selected six datasets from diverse categories: Beauty, Clothing, Sport, Toys, Kitchen, and Phone. In these datasets, each review rating signifies a positive user--item interaction. Following the standard practice in prior research \cite{he2016vbpr,he2020lightgcn,zhang2021mining} and to facilitate fair comparison with existing methods, we applied core-5 filtering, which refines the dataset ensuring each user and item has a minimum of five interactions. 
Dataset details are presented in Table~\ref{tab:datasets}.

\begin{table}[h]
\footnotesize
\setlength{\abovecaptionskip}{-0.2cm}
\begin{tabular}{l|c|c|c|c|c|c}
\hline
 & \textbf{Beauty} & \textbf{Clothing} & \textbf{Sports} & \textbf{Toys} & \textbf{Kitchen} & \textbf{Phone} \\ \hline
\textbf{User \#}  & 22,363  & 39,387 & 35,598 & 19,412 & 27,879 & 66,519 \\ \hline
\textbf{Item \#}  & {12,101} & {23,033} & {18,357} & {11,924}  & {10,429}  & {28,237}   \\ \hline
\textbf{Inter. \#}  & {198,502} & {278,677} & {296,337} & {167,597} & {194,439} & {551,682}  \\ \hline
\textbf{Avg Len. \#}  & {8.88}   & {7.12}     & {8.46}   & {8.79} & {7.19}    & {8.35}  \\ \hline
\textbf{Sparsity}  & {99.93\%}   & {99.97\%}     & {99.95\%}   & {99.93\%} & {99.93\%}    & {99.97\%}  \\ \hline
\end{tabular}
\caption{Dataset Statistics after preprocessing. }
\label{tab:datasets}
\end{table}

\textit{Baselines.}
We compare against  three groups of models.
(A) \textbf{Basic SR models} include \textit{GRU4Rec}~\cite{hidasi2015session} using Gated Recurrent Units (GRU) to model the sequential dependencies between items; \textit{SASRec}~\cite{kang2018self} employing a self-attention mechanism to capture long-term dependencies more effectively; and \textit{SR-GNN}~\cite{wu2019session}, a graph-based approach, incorporating both user--item interactions and item--item relationships to capture higher-order dependencies in sequential data. 
(B) \textbf{Multi-modal collaborative models} include 
\textit{MGAT}~\cite{tao2020mgat} focusing on disentangling personal interests by modality. It employs a graph attention network to integrate information from different modalities; \textit{MMGCN}~\cite{wei2019mmgcn} integrating multimodal features into a graph-based framework. It utilizes a message-passing scheme to learn the representations of users and items; \textit{BM3}~\cite{zhou2023bootstrap} bootstraping latent contrastive views of user/item representations, optimizing multimodal objectives for learning.
(C) \textbf{Feature-enriched SR models} include \textit{NOVA}~\cite{liu2021noninvasive} and \textit{DIF-SR}~\cite{xie2022decoupled} as state-of-the-art non-invasive fusion methods; \textit{Trans2D}~\cite{singer2022sequential} as holistic fusion methods.
We also used modified versions known as GRU4Rec$^{F}$ (late fusion) and SASRec$^{F}$ (early fusion), based on the GRU4Rec and SASRec models. These determine the best fusion choice for each, as seen in the intro case study.

\textit{Evaluation Protocol}
We follow convention and split each user's sequence into training and test datasets. Specifically, the last 20\% of the sequence is used as the test dataset, and the remaining 80\%, training. 
By pre-filtering sequences with a length of less than 5, we ensure that every user has at least one data point included in the test set.
We utilize two commonly-used ranking-based evaluation metrics, hit ratio (HR) and mean reciprocal rank (MRR), to assess performance. Higher values of HR and MRR indicate better model performance.

\textit{Parameter Settings.}
We standardize the embedding dimension to 128 and the batch size to 512, for all models. For hyperparameter tuning, we include learning rates ranging from \{1e-1, 1e-2, 1e-3, 1e-4\}, $L_2$ regularization values from \{0, 1e-1, 1e-2, 1e-3, 1e-4, 1e-5\}, and dropout ratios spanning from 0 to 0.9. We employ Adam optimization \cite{kingma2014adam}.
In addition, we experimented with layer sizes of \{1,2,3,4\} for the graph aggregator. , 
We report average performances from 5 repetitions of each experiment.

\begin{table*}[t]
\renewcommand{\arraystretch}{0.9}
\small
\begin{tabular}{c|l|ccc|ccc|ccccc|c}
\toprule
    & Metric  & GRU4Rec & SASRec & SR-GNN & MMGCN & MGAT & BM3 & GRU4Rec$^F$ & SASRec$^F$ & NOVA & DIF-SR & Trans2D & MMSR   \\ \hline \midrule
\multirow{4}{*}{\rotatebox{90}{\textbf{Beauty}}}   
     & HR@5 & 5.6420 & 6.1900 & 4.1483           & 2.6534 & 4.0870 & 4.8713           & 3.7682 & 6.4021 & 4.2219 & \underline{6.5789} & 6.0191          & \textbf{7.1563$^{*}$} \\
     & MRR@5 & 3.1110 & 3.2165 & 2.2123           & 1.2534 & 2.0297 & 2.3349          & 2.0793 & 3.7990 & 2.1785 & \underline{4.0735} & 3.4387          & \textbf{4.4429$^{*}$} \\
     & HR@20 & 12.7217 & 14.0681 & 10.2351       & 7.0443 & 9.1126 & 10.2640       & 9.4868 & 14.0269 & 10.7978 & \underline{14.0137} & 13.2214         & \textbf{14.1470$^{*}$} \\
     & MRR@20 & 3.7714 & 3.9668 & 2.7911           & 1.5263 & 2.6714 & 3.1945         & 2.6006 & 4.5073 & 2.8160 & \underline{4.7983} & 3.9460          & \textbf{5.0433$^{*}$} \\
\midrule
\multirow{4}{*}{\rotatebox{90}{\textbf{Clothing}}}     
     & HR@5 & 1.3340 & 1.5885 & 0.8547           & 0.5231 & 0.9613 & 1.2851           & 0.9501 & \underline{1.8430} & 1.2937 & 1.5524 & 1.3929          & \textbf{1.8684$^{*}$} \\
     & MRR@5 & 0.6765 & 0.7820 & 0.4555           & 0.2128 & 0.5470 & 0.5460          & 0.5212 & \underline{0.9470} & 0.6503 & 0.7961 & 0.6682          & \textbf{1.1365$^{*}$} \\
     & HR@20 & 3.8111 & 3.9574 & 2.7528           & 1.7847 & 2.7363 & 3.5072          & 2.8610 & \underline{4.2048} & 3.4866 & 4.0571 & 4.0683          & \textbf{4.4136$^{*}$} \\
     & MRR@20 & 0.9418 & 1.0339 & 0.6251           & 0.4359 & 0.7548 & 0.9045         & 0.6955 & \underline{1.2814} & 0.8783 & 1.0530 & 1.0391          & \textbf{1.3344$^{*}$} \\
\midrule
\multirow{4}{*}{\rotatebox{90}{\textbf{Sport}}}     
     & HR@5 & 2.4388 & 2.9549 & 2.0742           & 1.2020 & 2.0418 & 2.3096           & 1.8929 & \underline{3.1063} & 2.1539 & 2.5145 & 2.7168          & \textbf{3.2657$^{*}$} \\
     & MRR@5 & 1.2696 & 1.5858 & 1.0790           & 0.5688 & 0.8762 & 0.9963          & 0.9786 & \underline{1.6997} & 1.1271 & 1.3469 & 1.4235          & \textbf{1.9846$^{*}$} \\
     & HR@20 & 6.6430 & 7.2208 & 5.4376           & 3.6492 & 5.2197 & 5.3184          & 5.4834 & \underline{7.3683} & 5.8062 & 7.0774 & 6.9453          & \textbf{7.7466$^{*}$} \\
     & MRR@20 & 1.6947 & 2.0357 & 1.4349           & 0.8645 & 1.3002 & 1.5245         & 1.3274 & \underline{2.1427} & 1.5648 & 1.9214 & 1.7058          & \textbf{2.2826$^{*}$} \\
\midrule
\multirow{4}{*}{\rotatebox{90}{\textbf{Toys}}}    
    & HR@5 & 3.8663 & 5.0902 & 2.7329           & 1.7592 & 2.3746 & 3.9084            & 2.1974 & 5.2328 & 3.7899 & \underline{5.2363} & 4.1908          & \textbf{6.1159$^{*}$} \\
    & MRR@5 & 2.0022 & 2.7536 & 1.4878           & 0.7869 & 1.1369 & 2.0352           & 1.1576 & 3.0801 & 1.9641 & \underline{3.1944} & 2.2370          & \textbf{3.8987$^{*}$} \\
    & HR@20 & 10.0727 & 11.8668 & 6.7452        & 4.5497 & 5.9223 & 8.7071         & 6.0638 & 11.7485 & 9.0609 & \underline{12.0284} & 10.5082          & \textbf{12.1192$^{*}$} \\
    & MRR@20 & 2.7267 & 3.4228 & 1.8655           & 1.1256 & 1.5314 & 2.5623          & 1.5230 & 3.6812 & 2.4502 & \underline{3.8777} & 2.9298          & \textbf{4.3551$^{*}$} \\
\midrule
\multirow{4}{*}{\rotatebox{90}{\textbf{Kitchen}}}  
    & HR@5 & 1.1759 & 1.8012 & 1.1024           & 0.6671 & 1.2225 & 1.4399            & 1.1323 & \underline{1.9077} & 1.2558 & 1.5828 & 1.3463          & \textbf{2.2145$^{*}$} \\
    & MRR@5 & 0.5824 & 0.9729 & 0.5877           & 0.3154 & 0.4882 & 0.7012           & 0.5586 & \underline{1.1268} & 0.6279 & 0.8499 & 0.7413          & \textbf{1.4238$^{*}$} \\
    & HR@20 & 3.5640 & 4.2021 & 3.3255           & 2.2404 & 3.5206 & 3.4157           & 3.5449 & \underline{4.3187} & 3.5332 & 4.2766 & 3.8158          & \textbf{4.4535$^{*}$} \\
    & MRR@20 & 0.8277 & 1.2043 & 0.8507           & 0.5210 & 0.6898 & 0.8832          & 0.7817 & \underline{1.3862} & 0.8349 & 1.1041 & 0.8682          & \textbf{1.6086$^{*}$} \\
\midrule
\multirow{4}{*}{\rotatebox{90}{\textbf{Phone}}}    
    & HR@5 & 5.6626 & 6.4435 & 5.3128           & 3.2823 & 4.4046 & 4.9338            & 4.1188 & \underline{6.6908} & 5.3581 & 6.0666 & 6.0646          & \textbf{6.9550$^{*}$} \\
    & MRR@5 & 2.8765 & 3.4998 & 2.7221           & 1.4397 & 1.8735 & 2.3515           & 2.0211 & \underline{3.6643} & 2.7899 & 3.2383 & 3.0125          & \textbf{3.9911$^{*}$} \\
    & HR@20 & 13.4539 & 14.1525 & 12.1363      & 8.3255 & 10.9956 & 11.0081        & 11.3945 & \underline{14.6771} & 12.3232 & 14.6781 & 13.8446        & \textbf{14.9509$^{*}$} \\
    & MRR@20 & 3.7002 & 4.3182 & 3.4807           & 2.0647 & 3.0360 & 3.2278          & 3.0653 & \underline{4.5001} & 3.5063 & 4.2540 & 3.8798          & \textbf{4.5747$^{*}$} \\
\bottomrule
\end{tabular}
\caption{Overall Performance (\%). Bold ones indicate the best performances, while underlined ones indicate the best among baselines. $*$ indicates a statistically significant level $p$-value$<0.05$ comparing MMSR with the best baseline.}
\vspace{-1.5em}
\label{tab:result}
\end{table*}

\subsection{Overall Performance}

MMSR consistently outperforms other models (Table~\ref{tab:result}). It shows a significant improvement in HR (8.6\% for Top-5, 2.8\% for Top-20) and MRR (17.2\% for Top-5, 7.6\% for Top-20), on average. Our approach of fusing modal features enhances recommendation precision, ranking preferred items higher.

Comparing the basic sequential recommendation baseline with our baseline that includes modalities as side features, the latter is stronger overall. SASRec stands out among the baseline models, demonstrating the excellent performance of attention in sequential recommendation. In contrast, SR-GNN, the existing graph-based baseline, performs poorly, highlighting the superiority of our method in utilizing the graph.
Among the sequence recommendation baselines enhanced with modal features, DIF-SR and SASRec$^{F}$ perform best, demonstrating that attention effectively enhances early fusion (both invasive and non-invasive).
SASRec$^{F}$ adopts an invasive early fusion approach, directly fusing modal representation into item representation. In contrast, DIF-SR uses a non-invasive approach, where modal features are not fully integrated into the item representation vector. 
However, contrary to previous findings \cite{liu2021noninvasive}, our analysis shows that the invasive approach can be comparatively effective. 
This can be attributed to our modality codes (from the autoencoder), which introduce a more generalized modality representation for items, instead of too specific representation.

Existing multi-modal recommendation baselines focusing on inter-modality modeling with collaborative signals (MGCN, MGNN, BM3) do not incorporate sequential relationships, resulting in poor performance. 
It reveals that, for the SR task, besides inter-modality relationships, considering the intra-modality sequential relationships remains vital.
Our proposed method fills this gap and is necessary for improving sequence recommendation tasks.

\begin{table}[t]
\setlength{\abovecaptionskip}{-0.1cm}
\small
\tabcolsep=1.25mm
\begin{tabular}{l|cc|cc|cc}
\toprule
\multirow{2}{*}{Model} & \multicolumn{2}{|c}{\textbf{Beauty}} & \multicolumn{2}{|c}{\textbf{Clothing}} & \multicolumn{2}{|c}{\textbf{Sport}}  \\
  & HR@5 & MRR@5 & HR@5 & MRR@5 & HR@5 & MRR@5    \\
\midrule
\textit{GCN} & 5.6348 & 3.163 & 1.2340 & 0.6465 & 2.3177 & 1.1424 \\
\textit{GraphSAGE} & 5.5773 & 3.1283 & 1.3801 & 0.8552 & 2.2496 & 1.3473 \\
\textit{GAT} & 5.7116 & 3.1941 & 1.4092 & 0.8332 & 2.3452 & 1.3825 \\
\textit{Graphormer} & 5.9267 & 3.3029 & 1.4573 & 0.9029 & 2.3069 & 1.3756 \\
\midrule
\textit{RGAT} & 6.8157 & 3.9783 & 1.7352 & 1.0873 & 2.8609 & 1.7133 \\
\textit{HGNN} & 6.9701 & 4.1276 & 1.7721 & 1.1084 & 2.9682 & 1.7776 \\
\textit{HGAT} & 7.0671 & 4.2494 & 1.8448 & 1.1417 & 3.0458 & 1.8501 \\
\midrule
\textbf{HAN-GNN} & 7.1386 & 4.6244 & 2.0402 & 1.2642 & 3.3255 & 1.9916 \\
\bottomrule
\end{tabular}
\caption{Graph Aggregator Comparison.}
\vspace{-1.5em}
\label{tab:aggregator}
\end{table}

\subsection{Graph Aggregator Study}
\label{ss:gas}
In our paper, we designed a graph neural network specifically for integrating multi-modal features. To demonstrate its superiority over other graph neural networks, we compared it against several popular models, including GCN, GraphSAGE, GAT, and Graphormer, which do not consider heterogeneity; as well as RGAT, which considers heterogeneity in edge types; and HGNN and HGAT, which consider heterogeneity in node types.
Table~\ref{tab:aggregator} shows that our HAN-GNN method consistently outperforms other approaches. 
When comparing GAT and Graphormer, incorporating Transformer structures into graph neural networks is effective 
over traditional content-based attention. In MSGraph, incorporating heterogeneity in modality-enriched graphs leads to significant performance improvements compared to models that do not consider heterogeneity 
Further comparing HGNN and RGAT, we find that the heterogeneity of nodes is more important, particularly in distinguishing modality information from item node information. Thus, our non-invasive approach is more effective in handling heterogeneous information. 

\subsection{Ablation Study}
\label{ss:as}
To better understand the superiority of our approach, we conducted an ablation study on HAN-GNN.
In Table~\ref{tab:ablation}, $ho$ and $he$ signify HAN-GNN propagation solely through homogeneous or heterogeneous relations, respectively.
$hohe$ signifies the use of Homo--Hetero Ordering Fusion, while $heho$ represents Hetero--Homo Ordering Fusion. 
\textit{``NI''} signifies the non-invasive fusion ordering for each of them. 
\textit{``Synchronous''} refers to Equation~\ref{eq:asychronous}, which simply concatenates and linearly transforms homogeneous and heterogeneous information.

Examining the fusion of $ho$ and $he$ only, we found that the Sport dataset performs better when considering homogeneous relationships, while the Beauty and Clothing datasets benefit more from considering only heterogeneous information. This suggests that users in the latter scenarios rely more on either visual or textual information for ordering decisions, while this is not the case in the Sport dataset. 
Regarding fusion order, for invasive fusion, fusing homogeneous information before heterogeneous information ($hohe$) consistently yields better performance, comparing $heho$. However, for non-invasive fusion, the difference between order $NI(hohe)$ and $NI(heho)$ is not significant. This suggests that under invasive fusion, early fusion of heterogeneous attributes may disrupt the original item representation; but that non-invasive fusion alleviates this issue.
Furthermore, considering both fusion orders simultaneously (\textit{synchronous} fusion) does not perform as well as each order separately. However, our asynchronous update method (final HAN-GNN model) significantly improves performance compared to considering each order separately. In another words, our HAN-GNN model outperforms both fusion orders individually.

We also find removing either position embedding $\mathbf{e}^{po}$ or node type embedding $\mathbf{e}^{ty}$ in the representation stage noticeably deteriorates performance, validating the importance of retaining sequence and node type information in graph approaches.

\begin{table}[t]
\small
\tabcolsep=1.2mm
\begin{tabular}{l|cc|cc|cc}
\toprule
\multirow{2}{*}{Model} & \multicolumn{2}{|c}{\textbf{Beauty}} & \multicolumn{2}{|c}{\textbf{Clothing}} & \multicolumn{2}{|c}{\textbf{Sport}}  \\
 & HR@5 & MRR@5 & HR@5 & MRR@5 & HR@5 & MRR@5    \\
\midrule
\textbf{HAN-GNN} & 7.1386 & 4.6244 & 2.0402 & 1.2642 & 3.3255 & 1.9916 \\
\textit{Synchronous} & 6.8912 & 4.4515 & 1.7857 & 1.0681 & 3.0924 & 1.7849 \\
\textit{$NI(hohe)$} & 6.8900 & 4.4616 & 1.9999 & 1.2357 & 3.0616 & 1.8792 \\
\textit{$NI(heho)$} & 6.8897 & 4.5528 & 1.3932 & 0.7655 & 3.0087	& 1.7051  \\
\textit{$hohe$} & 6.8971 & 4.4245 & 1.9575 & 1.2169 & 3.0565 & 1.8793 \\
\textit{$heho$} & 6.5406 & 4.3117 & 1.1495 & 0.6398 & 2.8871 & 1.6654 \\
$ho$ & 6.6702 & 4.1004 & 1.6012 & 0.9069 & 3.0306 & 1.7412 \\
$he$ & 6.9354 & 4.446 & 1.9957 & 1.2236 & 3.0047 & 1.8648 \\
\midrule 
\textit{w/o $\mathbf{e}^{po}$} & 6.9664 & 4.5653 & 2.0665 & 1.2581 & 3.1547 & 1.8968 \\
\textit{w/o $\mathbf{e}^{ty}$} & 6.9390 & 4.5074 & 2.0370 & 1.2593 & 3.2112 & 1.9854 \\
\bottomrule
\end{tabular}
\caption{Ablation analysis, evaluated with (HR, MRR)@5. The relation ablation is based on a GCN aggregator.}
\vspace{-1.5em}
\label{tab:ablation}
\end{table}

\subsection{Robustness to Missing Modalities}

\begin{figure}[t]
    \centering
    \setlength{\abovecaptionskip}{-0cm}
    \setlength{\belowcaptionskip}{-0.3cm}
    \includegraphics[width=0.48\textwidth]{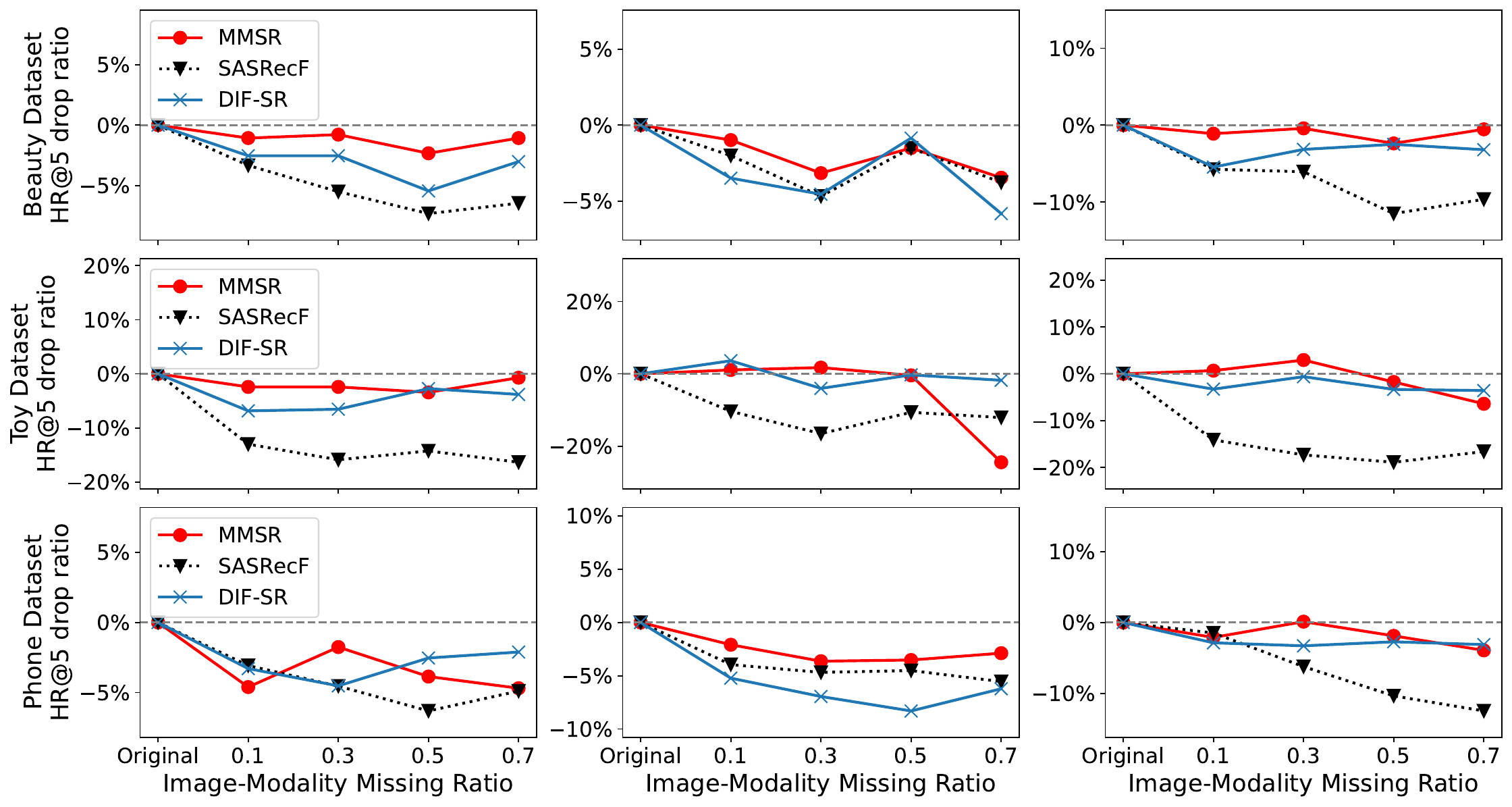}
  \caption{MMSR robustness against missing modalities.}
    \label{fig:robustness}
\end{figure}


Missing modalities are a common issue in real-world applications, and the traditional approach of filling missing features with default values is 
fragile.
Our method addresses this by utilizing graphs, which naturally handle missing modality nodes. Instead of replacing them with defaults, we 
simply
remove such nodes from the graph. We also incorporate global attention during node aggregation to ensure that modality-specific item nodes are aware of relevant modal nodes in the sequence.

In Figure~\ref{fig:robustness}, we compare the robustness of our method (MMSR) with the best-performing baselines, SASRecF and DIF-SR. The ``\textit{Image}''/``\textit{Text}''/``\textit{Mix}'' indicates the percentage of missing image features, text features, or both. We selected a missing ratio ($e$) between $0.1$ and $0.7$ for analysis.
MMSR shows robustness in scenarios with missing modalities (with $e$ in $0.1\sim0.5$), even achieving improvements under certain degrees of missing modalities.
\reedit{
This is akin to adversarial training \cite{he2018adversarial} where the introduction of a low level of noise enhances performance.
When significant modality information is lost ($e=0.7$), all methods show a substantial performance drop, highlighting the critical role of modality features.
}
For mixed missing modalities, MMSR is consistently more stable than other approaches. However, for text missing in Toy dataset and image missing in Phone dataset, MMSR's stability varies. This suggests that text and image nodes are more important modalities -- phones with comparable designs or toys with analogous textual descriptions indicate stronger associations -- respectively, in these datasets. 


\subsection{Modality-enriched Graph Construction}

Constructing a graph from a user's historical sequence can be challenging, as having too many modality nodes can result in an overly sparse graph. We thus compare different settings within our graph construction method (using a modality code set and soft links between original modalities and modality codes to improve the graph density).
The x-axis represents the cluster number (i.e., the number of modality codes), while the y-axis represents the number of codes corresponding to an original modality (i.e., $k$).

\begin{figure}[t]
    \centering
    \setlength{\abovecaptionskip}{-0cm}
    \setlength{\belowcaptionskip}{-0.6cm}
    \includegraphics[width=0.48\textwidth]{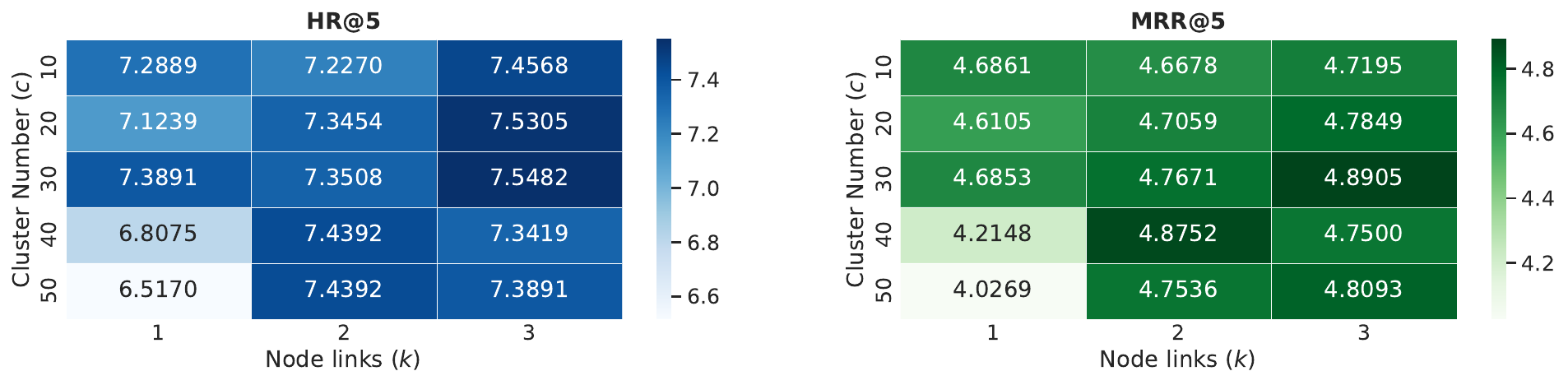}
  \caption{
  The performance comparison with different MSGraph construction parameters on the Beauty dataset.
  }
  \label{fig:c_k}
\end{figure}

We see that using modality codes achieves better performance 
over not using modality codes (compare \textit{HR@5}: $7.4263$, \textit{MRR@5}: $4.7469$ to results in Figure~\ref{fig:c_k}). 
Secondly, we observed that a larger value of $c$ does not necessarily lead to better performance, as the optimal point is typically between 20 and 30. As $c$ increases, the optimal value of $k$ increases accordingly.
Finally, the utilization of modality codes is consistent with the findings of previous studies \cite{hou2023learning,rajput2018recommender}, which demonstrated their positive impact on performance.

\section{Conclusion and Future Work}
We introduce a Multi-Modality enriched Sequential
Recommendation framework while optimally fuses modality features in sequential recommendation. 
Our approach tackles the complexity of fusing multi-modalities in sequential tasks, where fusion order notably influences the recommendation model performance.
To drive MMSR, we develop a novel graph aggregation mechanism
(HAN-GNN) that employs a dual graph attention network and asynchronous updating strategy. HAN-GNN flexibly integrates  modality information while preserving sequential relationships.
MMSR consistently outperforms state-of-the-art baselines, even under challenging missing-modality scenarios.  This makes it a flexible and robust solution for real-world applications.

MMSR is easily extensible, allowing for 
expansion to 
additional modalities.
We are optimistic about its utility in industrial contexts. 
Furthermore, exploring the interpretability of complex modal relationships in modality-enriched SR opens up new horizons for future research. Unraveling how and when sequentiality or interdependent relationships become pivotal 
could lead to more nuanced and efficient recommendation.

\begin{acks}
We thank the new deep learning computing framework MindSpore~\cite{mindspore2020mindspore} for the partial support of this work.
\end{acks}

\bibliographystyle{ACM-Reference-Format}
\bibliography{sample-base}


\begin{thebibliography}{57}


\ifx \showCODEN    \undefined \def \showCODEN     #1{\unskip}     \fi
\ifx \showDOI      \undefined \def \showDOI       #1{#1}\fi
\ifx \showISBNx    \undefined \def \showISBNx     #1{\unskip}     \fi
\ifx \showISBNxiii \undefined \def \showISBNxiii  #1{\unskip}     \fi
\ifx \showISSN     \undefined \def \showISSN      #1{\unskip}     \fi
\ifx \showLCCN     \undefined \def \showLCCN      #1{\unskip}     \fi
\ifx \shownote     \undefined \def \shownote      #1{#1}          \fi
\ifx \showarticletitle \undefined \def \showarticletitle #1{#1}   \fi
\ifx \showURL      \undefined \def \showURL       {\relax}        \fi
\providecommand\bibfield[2]{#2}
\providecommand\bibinfo[2]{#2}
\providecommand\natexlab[1]{#1}
\providecommand\showeprint[2][]{arXiv:#2}

\bibitem[min(2020)]%
        {mindspore2020mindspore}
 \bibinfo{year}{2020}\natexlab{}.
\newblock \showarticletitle{MindSpore}.
\newblock \bibinfo{journal}{\emph{https://www.mindspore.cn}}.
\newblock


\bibitem[Baldi and Hornik(1989)]%
        {baldi1989neural}
\bibfield{author}{\bibinfo{person}{Pierre Baldi} {and} \bibinfo{person}{Kurt
  Hornik}.} \bibinfo{year}{1989}\natexlab{}.
\newblock \showarticletitle{Neural networks and principal component analysis:
  Learning from examples without local minima}.
\newblock \bibinfo{journal}{\emph{Neural networks}} \bibinfo{volume}{2},
  \bibinfo{number}{1} (\bibinfo{year}{1989}), \bibinfo{pages}{53--58}.
\newblock


\bibitem[Chen et~al\mbox{.}(2022)]%
        {chen2022breaking}
\bibfield{author}{\bibinfo{person}{Feiyu Chen}, \bibinfo{person}{Junjie Wang},
  \bibinfo{person}{Yinwei Wei}, \bibinfo{person}{Hai-Tao Zheng}, {and}
  \bibinfo{person}{Jie Shao}.} \bibinfo{year}{2022}\natexlab{}.
\newblock \showarticletitle{Breaking Isolation: Multimodal Graph Fusion for
  Multimedia Recommendation by Edge-wise Modulation}. In
  \bibinfo{booktitle}{\emph{Proceedings of the 30th ACM International
  Conference on Multimedia}}. \bibinfo{pages}{385--394}.
\newblock


\bibitem[Chen and Wong(2020)]%
        {chen2020handling}
\bibfield{author}{\bibinfo{person}{Tianwen Chen} {and} \bibinfo{person}{Raymond
  Chi-Wing Wong}.} \bibinfo{year}{2020}\natexlab{}.
\newblock \showarticletitle{Handling information loss of graph neural networks
  for session-based recommendation}. In \bibinfo{booktitle}{\emph{Proceedings
  of the 26th ACM SIGKDD International Conference on Knowledge Discovery \&
  Data Mining}}. \bibinfo{pages}{1172--1180}.
\newblock


\bibitem[Chi et~al\mbox{.}(2016)]%
        {chi2016ubishop}
\bibfield{author}{\bibinfo{person}{Heng-Yu Chi}, \bibinfo{person}{Chun-Chieh
  Chen}, \bibinfo{person}{Wen-Huang Cheng}, {and} \bibinfo{person}{Ming-Syan
  Chen}.} \bibinfo{year}{2016}\natexlab{}.
\newblock \showarticletitle{UbiShop: Commercial item recommendation using
  visual part-based object representation}.
\newblock \bibinfo{journal}{\emph{Multimedia Tools and Applications}}
  \bibinfo{volume}{75} (\bibinfo{year}{2016}), \bibinfo{pages}{16093--16115}.
\newblock


\bibitem[Deldjoo et~al\mbox{.}(2018)]%
        {deldjoo2018audio}
\bibfield{author}{\bibinfo{person}{Yashar Deldjoo},
  \bibinfo{person}{Mihai~Gabriel Constantin}, \bibinfo{person}{Hamid
  Eghbal-Zadeh}, \bibinfo{person}{Bogdan Ionescu}, \bibinfo{person}{Markus
  Schedl}, {and} \bibinfo{person}{Paolo Cremonesi}.}
  \bibinfo{year}{2018}\natexlab{}.
\newblock \showarticletitle{Audio-visual encoding of multimedia content for
  enhancing movie recommendations}. In \bibinfo{booktitle}{\emph{Proceedings of
  the 12th ACM Conference on Recommender Systems}}. \bibinfo{pages}{455--459}.
\newblock


\bibitem[Deldjoo et~al\mbox{.}(2021)]%
        {deldjoo2021study}
\bibfield{author}{\bibinfo{person}{Yashar Deldjoo}, \bibinfo{person}{Tommaso
  Di~Noia}, \bibinfo{person}{Daniele Malitesta}, {and}
  \bibinfo{person}{Felice~Antonio Merra}.} \bibinfo{year}{2021}\natexlab{}.
\newblock \showarticletitle{A study on the relative importance of convolutional
  neural networks in visually-aware recommender systems}. In
  \bibinfo{booktitle}{\emph{Proceedings of the IEEE/CVF Conference on Computer
  Vision and Pattern Recognition}}. \bibinfo{pages}{3961--3967}.
\newblock


\bibitem[Deng et~al\mbox{.}(2009)]%
        {deng2009imagenet}
\bibfield{author}{\bibinfo{person}{Jia Deng}, \bibinfo{person}{Wei Dong},
  \bibinfo{person}{Richard Socher}, \bibinfo{person}{Li-Jia Li},
  \bibinfo{person}{Kai Li}, {and} \bibinfo{person}{Li Fei-Fei}.}
  \bibinfo{year}{2009}\natexlab{}.
\newblock \showarticletitle{Imagenet: A large-scale hierarchical image
  database}. In \bibinfo{booktitle}{\emph{2009 IEEE conference on computer
  vision and pattern recognition}}. Ieee, \bibinfo{pages}{248--255}.
\newblock


\bibitem[Devlin et~al\mbox{.}(2018)]%
        {devlin2018bert}
\bibfield{author}{\bibinfo{person}{Jacob Devlin}, \bibinfo{person}{Ming-Wei
  Chang}, \bibinfo{person}{Kenton Lee}, {and} \bibinfo{person}{Kristina
  Toutanova}.} \bibinfo{year}{2018}\natexlab{}.
\newblock \showarticletitle{Bert: Pre-training of deep bidirectional
  transformers for language understanding}.
\newblock \bibinfo{journal}{\emph{arXiv preprint arXiv:1810.04805}}
  (\bibinfo{year}{2018}).
\newblock


\bibitem[Geng et~al\mbox{.}(2015)]%
        {geng2015learning}
\bibfield{author}{\bibinfo{person}{Xue Geng}, \bibinfo{person}{Hanwang Zhang},
  \bibinfo{person}{Jingwen Bian}, {and} \bibinfo{person}{Tat-Seng Chua}.}
  \bibinfo{year}{2015}\natexlab{}.
\newblock \showarticletitle{Learning image and user features for recommendation
  in social networks}. In \bibinfo{booktitle}{\emph{Proceedings of the IEEE
  international conference on computer vision}}. \bibinfo{pages}{4274--4282}.
\newblock


\bibitem[Gong and Zhang(2016)]%
        {gong2016hashtag}
\bibfield{author}{\bibinfo{person}{Yuyun Gong} {and} \bibinfo{person}{Qi
  Zhang}.} \bibinfo{year}{2016}\natexlab{}.
\newblock \showarticletitle{Hashtag recommendation using attention-based
  convolutional neural network.}. In \bibinfo{booktitle}{\emph{IJCAI}}.
  \bibinfo{pages}{2782--2788}.
\newblock


\bibitem[Gu et~al\mbox{.}(2016)]%
        {gu2016iglasses}
\bibfield{author}{\bibinfo{person}{Xiaoling Gu}, \bibinfo{person}{Lidan Shou},
  \bibinfo{person}{Pai Peng}, \bibinfo{person}{Ke Chen}, \bibinfo{person}{Sai
  Wu}, {and} \bibinfo{person}{Gang Chen}.} \bibinfo{year}{2016}\natexlab{}.
\newblock \showarticletitle{iGlasses: A novel recommendation system for
  best-fit glasses}. In \bibinfo{booktitle}{\emph{Proceedings of the 39th
  International ACM SIGIR conference on research and development in information
  retrieval}}. \bibinfo{pages}{1109--1112}.
\newblock


\bibitem[He et~al\mbox{.}(2016)]%
        {he2016deep}
\bibfield{author}{\bibinfo{person}{Kaiming He}, \bibinfo{person}{Xiangyu
  Zhang}, \bibinfo{person}{Shaoqing Ren}, {and} \bibinfo{person}{Jian Sun}.}
  \bibinfo{year}{2016}\natexlab{}.
\newblock \showarticletitle{Deep residual learning for image recognition}. In
  \bibinfo{booktitle}{\emph{Proceedings of the IEEE conference on computer
  vision and pattern recognition}}. \bibinfo{pages}{770--778}.
\newblock


\bibitem[He and McAuley(2016a)]%
        {he2016ups}
\bibfield{author}{\bibinfo{person}{Ruining He} {and} \bibinfo{person}{Julian
  McAuley}.} \bibinfo{year}{2016}\natexlab{a}.
\newblock \showarticletitle{Ups and downs: Modeling the visual evolution of
  fashion trends with one-class collaborative filtering}. In
  \bibinfo{booktitle}{\emph{proceedings of the 25th international conference on
  world wide web}}. \bibinfo{pages}{507--517}.
\newblock


\bibitem[He and McAuley(2016b)]%
        {he2016vbpr}
\bibfield{author}{\bibinfo{person}{Ruining He} {and} \bibinfo{person}{Julian
  McAuley}.} \bibinfo{year}{2016}\natexlab{b}.
\newblock \showarticletitle{VBPR: visual bayesian personalized ranking from
  implicit feedback}. In \bibinfo{booktitle}{\emph{Proceedings of the AAAI
  conference on artificial intelligence}}, Vol.~\bibinfo{volume}{30}.
\newblock


\bibitem[He et~al\mbox{.}(2020)]%
        {he2020lightgcn}
\bibfield{author}{\bibinfo{person}{Xiangnan He}, \bibinfo{person}{Kuan Deng},
  \bibinfo{person}{Xiang Wang}, \bibinfo{person}{Yan Li},
  \bibinfo{person}{Yongdong Zhang}, {and} \bibinfo{person}{Meng Wang}.}
  \bibinfo{year}{2020}\natexlab{}.
\newblock \showarticletitle{Lightgcn: Simplifying and powering graph
  convolution network for recommendation}. In
  \bibinfo{booktitle}{\emph{Proceedings of the 43rd International ACM SIGIR
  conference on research and development in Information Retrieval}}.
  \bibinfo{pages}{639--648}.
\newblock


\bibitem[He et~al\mbox{.}(2018)]%
        {he2018adversarial}
\bibfield{author}{\bibinfo{person}{Xiangnan He}, \bibinfo{person}{Zhankui He},
  \bibinfo{person}{Xiaoyu Du}, {and} \bibinfo{person}{Tat-Seng Chua}.}
  \bibinfo{year}{2018}\natexlab{}.
\newblock \showarticletitle{Adversarial personalized ranking for
  recommendation}. In \bibinfo{booktitle}{\emph{The 41st International ACM
  SIGIR conference on research \& development in information retrieval}}.
  \bibinfo{pages}{355--364}.
\newblock


\bibitem[Hidasi et~al\mbox{.}(2015)]%
        {hidasi2015session}
\bibfield{author}{\bibinfo{person}{Bal{\'a}zs Hidasi},
  \bibinfo{person}{Alexandros Karatzoglou}, \bibinfo{person}{Linas Baltrunas},
  {and} \bibinfo{person}{Domonkos Tikk}.} \bibinfo{year}{2015}\natexlab{}.
\newblock \showarticletitle{Session-based recommendations with recurrent neural
  networks}.
\newblock \bibinfo{journal}{\emph{arXiv preprint arXiv:1511.06939}}
  (\bibinfo{year}{2015}).
\newblock


\bibitem[Hidasi et~al\mbox{.}(2016)]%
        {hidasi2016parallel}
\bibfield{author}{\bibinfo{person}{Bal{\'a}zs Hidasi}, \bibinfo{person}{Massimo
  Quadrana}, \bibinfo{person}{Alexandros Karatzoglou}, {and}
  \bibinfo{person}{Domonkos Tikk}.} \bibinfo{year}{2016}\natexlab{}.
\newblock \showarticletitle{Parallel recurrent neural network architectures for
  feature-rich session-based recommendations}. In
  \bibinfo{booktitle}{\emph{Proceedings of the 10th ACM conference on
  recommender systems}}. \bibinfo{pages}{241--248}.
\newblock


\bibitem[Hou et~al\mbox{.}(2023)]%
        {hou2023learning}
\bibfield{author}{\bibinfo{person}{Yupeng Hou}, \bibinfo{person}{Zhankui He},
  \bibinfo{person}{Julian McAuley}, {and} \bibinfo{person}{Wayne~Xin Zhao}.}
  \bibinfo{year}{2023}\natexlab{}.
\newblock \showarticletitle{Learning vector-quantized item representation for
  transferable sequential recommenders}. In
  \bibinfo{booktitle}{\emph{Proceedings of the ACM Web Conference 2023}}.
  \bibinfo{pages}{1162--1171}.
\newblock


\bibitem[Hu et~al\mbox{.}(2022)]%
        {hu2022modeling}
\bibfield{author}{\bibinfo{person}{Hengchang Hu}, \bibinfo{person}{Liangming
  Pan}, \bibinfo{person}{Yiding Ran}, {and} \bibinfo{person}{Min-Yen Kan}.}
  \bibinfo{year}{2022}\natexlab{}.
\newblock \showarticletitle{Modeling and Leveraging Prerequisite Context in
  Recommendation}.
\newblock \bibinfo{journal}{\emph{CARS@RecSys}} (\bibinfo{year}{2022}).
\newblock


\bibitem[Kang and McAuley(2018)]%
        {kang2018self}
\bibfield{author}{\bibinfo{person}{Wang-Cheng Kang} {and}
  \bibinfo{person}{Julian McAuley}.} \bibinfo{year}{2018}\natexlab{}.
\newblock \showarticletitle{Self-attentive sequential recommendation}. In
  \bibinfo{booktitle}{\emph{2018 IEEE international conference on data mining
  (ICDM)}}. IEEE, \bibinfo{pages}{197--206}.
\newblock


\bibitem[Kingma and Ba(2014)]%
        {kingma2014adam}
\bibfield{author}{\bibinfo{person}{Diederik~P Kingma} {and}
  \bibinfo{person}{Jimmy Ba}.} \bibinfo{year}{2014}\natexlab{}.
\newblock \showarticletitle{Adam: A method for stochastic optimization}.
\newblock \bibinfo{journal}{\emph{arXiv preprint arXiv:1412.6980}}
  (\bibinfo{year}{2014}).
\newblock


\bibitem[Kipf and Welling(2016)]%
        {kipf2016semi}
\bibfield{author}{\bibinfo{person}{Thomas~N Kipf} {and} \bibinfo{person}{Max
  Welling}.} \bibinfo{year}{2016}\natexlab{}.
\newblock \showarticletitle{Semi-supervised classification with graph
  convolutional networks}.
\newblock \bibinfo{journal}{\emph{arXiv preprint arXiv:1609.02907}}
  (\bibinfo{year}{2016}).
\newblock


\bibitem[Lei et~al\mbox{.}(2019)]%
        {lei2019tissa}
\bibfield{author}{\bibinfo{person}{Chenyi Lei}, \bibinfo{person}{Shouling Ji},
  {and} \bibinfo{person}{Zhao Li}.} \bibinfo{year}{2019}\natexlab{}.
\newblock \showarticletitle{Tissa: A time slice self-attention approach for
  modeling sequential user behaviors}. In \bibinfo{booktitle}{\emph{The World
  Wide Web Conference}}. \bibinfo{pages}{2964--2970}.
\newblock


\bibitem[Li et~al\mbox{.}(2020)]%
        {li2020hierarchical}
\bibfield{author}{\bibinfo{person}{Xingchen Li}, \bibinfo{person}{Xiang Wang},
  \bibinfo{person}{Xiangnan He}, \bibinfo{person}{Long Chen},
  \bibinfo{person}{Jun Xiao}, {and} \bibinfo{person}{Tat-Seng Chua}.}
  \bibinfo{year}{2020}\natexlab{}.
\newblock \showarticletitle{Hierarchical fashion graph network for personalized
  outfit recommendation}. In \bibinfo{booktitle}{\emph{Proceedings of the 43rd
  International ACM SIGIR Conference on Research and Development in Information
  Retrieval}}. \bibinfo{pages}{159--168}.
\newblock


\bibitem[Lin et~al\mbox{.}(2022)]%
        {lin2022sequential}
\bibfield{author}{\bibinfo{person}{Kun Lin}, \bibinfo{person}{Zhenlei Wang},
  \bibinfo{person}{Shiqi Shen}, \bibinfo{person}{Zhipeng Wang},
  \bibinfo{person}{Bo Chen}, {and} \bibinfo{person}{Xu Chen}.}
  \bibinfo{year}{2022}\natexlab{}.
\newblock \showarticletitle{Sequential Recommendation with Decomposed Item
  Feature Routing}. In \bibinfo{booktitle}{\emph{Proceedings of the ACM Web
  Conference 2022}}. \bibinfo{pages}{2288--2297}.
\newblock


\bibitem[Liu et~al\mbox{.}(2021)]%
        {liu2021noninvasive}
\bibfield{author}{\bibinfo{person}{Chang Liu}, \bibinfo{person}{Xiaoguang Li},
  \bibinfo{person}{Guohao Cai}, \bibinfo{person}{Zhenhua Dong},
  \bibinfo{person}{Hong Zhu}, {and} \bibinfo{person}{Lifeng Shang}.}
  \bibinfo{year}{2021}\natexlab{}.
\newblock \showarticletitle{Noninvasive self-attention for side information
  fusion in sequential recommendation}. In
  \bibinfo{booktitle}{\emph{Proceedings of the AAAI Conference on Artificial
  Intelligence}}, Vol.~\bibinfo{volume}{35}. \bibinfo{pages}{4249--4256}.
\newblock


\bibitem[Liu et~al\mbox{.}(2019)]%
        {liu2019user}
\bibfield{author}{\bibinfo{person}{Fan Liu}, \bibinfo{person}{Zhiyong Cheng},
  \bibinfo{person}{Changchang Sun}, \bibinfo{person}{Yinglong Wang},
  \bibinfo{person}{Liqiang Nie}, {and} \bibinfo{person}{Mohan Kankanhalli}.}
  \bibinfo{year}{2019}\natexlab{}.
\newblock \showarticletitle{User diverse preference modeling by multimodal
  attentive metric learning}. In \bibinfo{booktitle}{\emph{Proceedings of the
  27th ACM international conference on multimedia}}.
  \bibinfo{pages}{1526--1534}.
\newblock


\bibitem[Lloyd(1982)]%
        {lloyd1982least}
\bibfield{author}{\bibinfo{person}{Stuart Lloyd}.}
  \bibinfo{year}{1982}\natexlab{}.
\newblock \showarticletitle{Least squares quantization in PCM}.
\newblock \bibinfo{journal}{\emph{IEEE transactions on information theory}}
  \bibinfo{volume}{28}, \bibinfo{number}{2} (\bibinfo{year}{1982}),
  \bibinfo{pages}{129--137}.
\newblock


\bibitem[Luo et~al\mbox{.}(2008)]%
        {luo2008personalized}
\bibfield{author}{\bibinfo{person}{Hangzai Luo}, \bibinfo{person}{Jianping
  Fan}, {and} \bibinfo{person}{Daniel~A Keim}.}
  \bibinfo{year}{2008}\natexlab{}.
\newblock \showarticletitle{Personalized news video recommendation}. In
  \bibinfo{booktitle}{\emph{Proceedings of the 16th ACM international
  conference on Multimedia}}. \bibinfo{pages}{1001--1002}.
\newblock


\bibitem[Malkiel et~al\mbox{.}(2020)]%
        {malkiel2020recobert}
\bibfield{author}{\bibinfo{person}{Itzik Malkiel}, \bibinfo{person}{Oren
  Barkan}, \bibinfo{person}{Avi Caciularu}, \bibinfo{person}{Noam Razin},
  \bibinfo{person}{Ori Katz}, {and} \bibinfo{person}{Noam Koenigstein}.}
  \bibinfo{year}{2020}\natexlab{}.
\newblock \showarticletitle{RecoBERT: A catalog language model for text-based
  recommendations}.
\newblock \bibinfo{journal}{\emph{arXiv preprint arXiv:2009.13292}}
  (\bibinfo{year}{2020}).
\newblock


\bibitem[Niu et~al\mbox{.}(2018)]%
        {niu2018neural}
\bibfield{author}{\bibinfo{person}{Wei Niu}, \bibinfo{person}{James Caverlee},
  {and} \bibinfo{person}{Haokai Lu}.} \bibinfo{year}{2018}\natexlab{}.
\newblock \showarticletitle{Neural personalized ranking for image
  recommendation}. In \bibinfo{booktitle}{\emph{Proceedings of the eleventh ACM
  international conference on web search and data mining}}.
  \bibinfo{pages}{423--431}.
\newblock


\bibitem[Oramas et~al\mbox{.}(2016)]%
        {oramas2016sound}
\bibfield{author}{\bibinfo{person}{Sergio Oramas},
  \bibinfo{person}{Vito~Claudio Ostuni}, \bibinfo{person}{Tommaso~Di Noia},
  \bibinfo{person}{Xavier Serra}, {and} \bibinfo{person}{Eugenio~Di Sciascio}.}
  \bibinfo{year}{2016}\natexlab{}.
\newblock \showarticletitle{Sound and music recommendation with knowledge
  graphs}.
\newblock \bibinfo{journal}{\emph{ACM Transactions on Intelligent Systems and
  Technology (TIST)}} \bibinfo{volume}{8}, \bibinfo{number}{2}
  (\bibinfo{year}{2016}), \bibinfo{pages}{1--21}.
\newblock


\bibitem[Raffel et~al\mbox{.}(2020)]%
        {raffel2020exploring}
\bibfield{author}{\bibinfo{person}{Colin Raffel}, \bibinfo{person}{Noam
  Shazeer}, \bibinfo{person}{Adam Roberts}, \bibinfo{person}{Katherine Lee},
  \bibinfo{person}{Sharan Narang}, \bibinfo{person}{Michael Matena},
  \bibinfo{person}{Yanqi Zhou}, \bibinfo{person}{Wei Li}, {and}
  \bibinfo{person}{Peter~J Liu}.} \bibinfo{year}{2020}\natexlab{}.
\newblock \showarticletitle{Exploring the limits of transfer learning with a
  unified text-to-text transformer}.
\newblock \bibinfo{journal}{\emph{The Journal of Machine Learning Research}}
  \bibinfo{volume}{21}, \bibinfo{number}{1} (\bibinfo{year}{2020}),
  \bibinfo{pages}{5485--5551}.
\newblock


\bibitem[Rajput et~al\mbox{.}(2018)]%
        {rajput2018recommender}
\bibfield{author}{\bibinfo{person}{Shashank Rajput}, \bibinfo{person}{Nikhil
  Mehta}, \bibinfo{person}{Anima Singh}, \bibinfo{person}{Raghunandan
  Keshavan}, \bibinfo{person}{Trung Vu}, \bibinfo{person}{Lukasz Heldt},
  \bibinfo{person}{Lichan Hong}, \bibinfo{person}{Yi Tay},
  \bibinfo{person}{Vinh~Q Tran}, \bibinfo{person}{Jonah Samost},
  {et~al\mbox{.}}} \bibinfo{year}{2018}\natexlab{}.
\newblock \showarticletitle{Recommender Systems with Generative Retrieval}.
\newblock  (\bibinfo{year}{2018}).
\newblock


\bibitem[Ran et~al\mbox{.}(2022)]%
        {ran2022pm}
\bibfield{author}{\bibinfo{person}{Yiding Ran}, \bibinfo{person}{Hengchang Hu},
  {and} \bibinfo{person}{Min-Yen Kan}.} \bibinfo{year}{2022}\natexlab{}.
\newblock \showarticletitle{PM K-LightGCN: Optimizing for Accuracy and
  Popularity Match in Course Recommendation}. In
  \bibinfo{booktitle}{\emph{Workshop of Multi-Objective Recommender Systems
  (MORS’22), in conjunction with the 16th ACM Conference on Recommender
  Systems, RecSys}}, Vol.~\bibinfo{volume}{22}. \bibinfo{pages}{2022}.
\newblock


\bibitem[Rao et~al\mbox{.}(2013)]%
        {rao2013entity}
\bibfield{author}{\bibinfo{person}{Delip Rao}, \bibinfo{person}{Paul McNamee},
  {and} \bibinfo{person}{Mark Dredze}.} \bibinfo{year}{2013}\natexlab{}.
\newblock \showarticletitle{Entity linking: Finding extracted entities in a
  knowledge base}.
\newblock \bibinfo{journal}{\emph{Multi-source, multilingual information
  extraction and summarization}} (\bibinfo{year}{2013}),
  \bibinfo{pages}{93--115}.
\newblock


\bibitem[Rashed et~al\mbox{.}(2022)]%
        {rashed2022context}
\bibfield{author}{\bibinfo{person}{Ahmed Rashed}, \bibinfo{person}{Shereen
  Elsayed}, {and} \bibinfo{person}{Lars Schmidt-Thieme}.}
  \bibinfo{year}{2022}\natexlab{}.
\newblock \showarticletitle{Context and Attribute-Aware Sequential
  Recommendation via Cross-Attention}. In \bibinfo{booktitle}{\emph{Proceedings
  of the 16th ACM Conference on Recommender Systems}}. \bibinfo{pages}{71--80}.
\newblock


\bibitem[Singer et~al\mbox{.}(2022)]%
        {singer2022sequential}
\bibfield{author}{\bibinfo{person}{Uriel Singer}, \bibinfo{person}{Haggai
  Roitman}, \bibinfo{person}{Yotam Eshel}, \bibinfo{person}{Alexander Nus},
  \bibinfo{person}{Ido Guy}, \bibinfo{person}{Or Levi}, \bibinfo{person}{Idan
  Hasson}, {and} \bibinfo{person}{Eliyahu Kiperwasser}.}
  \bibinfo{year}{2022}\natexlab{}.
\newblock \showarticletitle{Sequential modeling with multiple attributes for
  watchlist recommendation in e-commerce}. In
  \bibinfo{booktitle}{\emph{Proceedings of the Fifteenth ACM International
  Conference on Web Search and Data Mining}}. \bibinfo{pages}{937--946}.
\newblock


\bibitem[Sun et~al\mbox{.}(2019)]%
        {sun2019bert4rec}
\bibfield{author}{\bibinfo{person}{Fei Sun}, \bibinfo{person}{Jun Liu},
  \bibinfo{person}{Jian Wu}, \bibinfo{person}{Changhua Pei},
  \bibinfo{person}{Xiao Lin}, \bibinfo{person}{Wenwu Ou}, {and}
  \bibinfo{person}{Peng Jiang}.} \bibinfo{year}{2019}\natexlab{}.
\newblock \showarticletitle{BERT4Rec: Sequential recommendation with
  bidirectional encoder representations from transformer}. In
  \bibinfo{booktitle}{\emph{Proceedings of the 28th ACM international
  conference on information and knowledge management}}.
  \bibinfo{pages}{1441--1450}.
\newblock


\bibitem[Tang and Wang(2018)]%
        {tang2018personalized}
\bibfield{author}{\bibinfo{person}{Jiaxi Tang} {and} \bibinfo{person}{Ke
  Wang}.} \bibinfo{year}{2018}\natexlab{}.
\newblock \showarticletitle{Personalized top-n sequential recommendation via
  convolutional sequence embedding}. In \bibinfo{booktitle}{\emph{Proceedings
  of the eleventh ACM international conference on web search and data mining}}.
  \bibinfo{pages}{565--573}.
\newblock


\bibitem[Tao et~al\mbox{.}(2020)]%
        {tao2020mgat}
\bibfield{author}{\bibinfo{person}{Zhulin Tao}, \bibinfo{person}{Yinwei Wei},
  \bibinfo{person}{Xiang Wang}, \bibinfo{person}{Xiangnan He},
  \bibinfo{person}{Xianglin Huang}, {and} \bibinfo{person}{Tat-Seng Chua}.}
  \bibinfo{year}{2020}\natexlab{}.
\newblock \showarticletitle{Mgat: Multimodal graph attention network for
  recommendation}.
\newblock \bibinfo{journal}{\emph{Information Processing \& Management}}
  \bibinfo{volume}{57}, \bibinfo{number}{5} (\bibinfo{year}{2020}),
  \bibinfo{pages}{102277}.
\newblock


\bibitem[Van~den Oord et~al\mbox{.}(2013)]%
        {van2013deep}
\bibfield{author}{\bibinfo{person}{Aaron Van~den Oord}, \bibinfo{person}{Sander
  Dieleman}, {and} \bibinfo{person}{Benjamin Schrauwen}.}
  \bibinfo{year}{2013}\natexlab{}.
\newblock \showarticletitle{Deep content-based music recommendation}.
\newblock \bibinfo{journal}{\emph{Advances in neural information processing
  systems}}  \bibinfo{volume}{26} (\bibinfo{year}{2013}).
\newblock


\bibitem[Velickovic et~al\mbox{.}(2017)]%
        {velickovic2017graph}
\bibfield{author}{\bibinfo{person}{Petar Velickovic}, \bibinfo{person}{Guillem
  Cucurull}, \bibinfo{person}{Arantxa Casanova}, \bibinfo{person}{Adriana
  Romero}, \bibinfo{person}{Pietro Lio}, \bibinfo{person}{Yoshua Bengio},
  {et~al\mbox{.}}} \bibinfo{year}{2017}\natexlab{}.
\newblock \showarticletitle{Graph attention networks}.
\newblock \bibinfo{journal}{\emph{stat}} \bibinfo{volume}{1050},
  \bibinfo{number}{20} (\bibinfo{year}{2017}), \bibinfo{pages}{10--48550}.
\newblock


\bibitem[Wei et~al\mbox{.}(2019)]%
        {wei2019mmgcn}
\bibfield{author}{\bibinfo{person}{Yinwei Wei}, \bibinfo{person}{Xiang Wang},
  \bibinfo{person}{Liqiang Nie}, \bibinfo{person}{Xiangnan He},
  \bibinfo{person}{Richang Hong}, {and} \bibinfo{person}{Tat-Seng Chua}.}
  \bibinfo{year}{2019}\natexlab{}.
\newblock \showarticletitle{MMGCN: Multi-modal graph convolution network for
  personalized recommendation of micro-video}. In
  \bibinfo{booktitle}{\emph{Proceedings of the 27th ACM international
  conference on multimedia}}. \bibinfo{pages}{1437--1445}.
\newblock


\bibitem[Wu et~al\mbox{.}(2022)]%
        {wu2022mm}
\bibfield{author}{\bibinfo{person}{Chuhan Wu}, \bibinfo{person}{Fangzhao Wu},
  \bibinfo{person}{Tao Qi}, \bibinfo{person}{Chao Zhang},
  \bibinfo{person}{Yongfeng Huang}, {and} \bibinfo{person}{Tong Xu}.}
  \bibinfo{year}{2022}\natexlab{}.
\newblock \showarticletitle{MM-Rec: Visiolinguistic Model Empowered Multimodal
  News Recommendation}. In \bibinfo{booktitle}{\emph{Proceedings of the 45th
  International ACM SIGIR Conference on Research and Development in Information
  Retrieval}}. \bibinfo{pages}{2560--2564}.
\newblock


\bibitem[Wu et~al\mbox{.}(2019)]%
        {wu2019session}
\bibfield{author}{\bibinfo{person}{Shu Wu}, \bibinfo{person}{Yuyuan Tang},
  \bibinfo{person}{Yanqiao Zhu}, \bibinfo{person}{Liang Wang},
  \bibinfo{person}{Xing Xie}, {and} \bibinfo{person}{Tieniu Tan}.}
  \bibinfo{year}{2019}\natexlab{}.
\newblock \showarticletitle{Session-based recommendation with graph neural
  networks}. In \bibinfo{booktitle}{\emph{Proceedings of the AAAI conference on
  artificial intelligence}}, Vol.~\bibinfo{volume}{33}.
  \bibinfo{pages}{346--353}.
\newblock


\bibitem[Xie et~al\mbox{.}(2022)]%
        {xie2022decoupled}
\bibfield{author}{\bibinfo{person}{Yueqi Xie}, \bibinfo{person}{Peilin Zhou},
  {and} \bibinfo{person}{Sunghun Kim}.} \bibinfo{year}{2022}\natexlab{}.
\newblock \showarticletitle{Decoupled side information fusion for sequential
  recommendation}. In \bibinfo{booktitle}{\emph{Proceedings of the 45th
  International ACM SIGIR Conference on Research and Development in Information
  Retrieval}}. \bibinfo{pages}{1611--1621}.
\newblock


\bibitem[Ying et~al\mbox{.}(2021)]%
        {ying2021transformers}
\bibfield{author}{\bibinfo{person}{Chengxuan Ying}, \bibinfo{person}{Tianle
  Cai}, \bibinfo{person}{Shengjie Luo}, \bibinfo{person}{Shuxin Zheng},
  \bibinfo{person}{Guolin Ke}, \bibinfo{person}{Di He},
  \bibinfo{person}{Yanming Shen}, {and} \bibinfo{person}{Tie-Yan Liu}.}
  \bibinfo{year}{2021}\natexlab{}.
\newblock \showarticletitle{Do transformers really perform badly for graph
  representation?}
\newblock \bibinfo{journal}{\emph{Advances in Neural Information Processing
  Systems}}  \bibinfo{volume}{34} (\bibinfo{year}{2021}),
  \bibinfo{pages}{28877--28888}.
\newblock


\bibitem[Yu et~al\mbox{.}(2014)]%
        {yu2014factor}
\bibfield{author}{\bibinfo{person}{Mo Yu}, \bibinfo{person}{Matthew Gormley},
  {and} \bibinfo{person}{Mark Dredze}.} \bibinfo{year}{2014}\natexlab{}.
\newblock \showarticletitle{Factor-based compositional embedding models}. In
  \bibinfo{booktitle}{\emph{NIPS Workshop on Learning Semantics}}.
  \bibinfo{pages}{95--101}.
\newblock


\bibitem[Yu et~al\mbox{.}(2018)]%
        {yu2018aesthetic}
\bibfield{author}{\bibinfo{person}{Wenhui Yu}, \bibinfo{person}{Huidi Zhang},
  \bibinfo{person}{Xiangnan He}, \bibinfo{person}{Xu Chen}, \bibinfo{person}{Li
  Xiong}, {and} \bibinfo{person}{Zheng Qin}.} \bibinfo{year}{2018}\natexlab{}.
\newblock \showarticletitle{Aesthetic-based clothing recommendation}. In
  \bibinfo{booktitle}{\emph{Proceedings of the 2018 world wide web
  conference}}. \bibinfo{pages}{649--658}.
\newblock


\bibitem[Zhang et~al\mbox{.}(2021)]%
        {zhang2021mining}
\bibfield{author}{\bibinfo{person}{Jinghao Zhang}, \bibinfo{person}{Yanqiao
  Zhu}, \bibinfo{person}{Qiang Liu}, \bibinfo{person}{Shu Wu},
  \bibinfo{person}{Shuhui Wang}, {and} \bibinfo{person}{Liang Wang}.}
  \bibinfo{year}{2021}\natexlab{}.
\newblock \showarticletitle{Mining latent structures for multimedia
  recommendation}. In \bibinfo{booktitle}{\emph{Proceedings of the 29th ACM
  International Conference on Multimedia}}. \bibinfo{pages}{3872--3880}.
\newblock


\bibitem[Zhang et~al\mbox{.}(2019)]%
        {zhang2019feature}
\bibfield{author}{\bibinfo{person}{Tingting Zhang}, \bibinfo{person}{Pengpeng
  Zhao}, \bibinfo{person}{Yanchi Liu}, \bibinfo{person}{Victor~S Sheng},
  \bibinfo{person}{Jiajie Xu}, \bibinfo{person}{Deqing Wang},
  \bibinfo{person}{Guanfeng Liu}, \bibinfo{person}{Xiaofang Zhou},
  {et~al\mbox{.}}} \bibinfo{year}{2019}\natexlab{}.
\newblock \showarticletitle{Feature-level Deeper Self-Attention Network for
  Sequential Recommendation.}. In \bibinfo{booktitle}{\emph{IJCAI}}.
  \bibinfo{pages}{4320--4326}.
\newblock


\bibitem[Zheng et~al\mbox{.}(2017)]%
        {zheng2017joint}
\bibfield{author}{\bibinfo{person}{Lei Zheng}, \bibinfo{person}{Vahid Noroozi},
  {and} \bibinfo{person}{Philip~S Yu}.} \bibinfo{year}{2017}\natexlab{}.
\newblock \showarticletitle{Joint deep modeling of users and items using
  reviews for recommendation}. In \bibinfo{booktitle}{\emph{Proceedings of the
  tenth ACM international conference on web search and data mining}}.
  \bibinfo{pages}{425--434}.
\newblock


\bibitem[Zhou et~al\mbox{.}(2023b)]%
        {zhou2023comprehensive}
\bibfield{author}{\bibinfo{person}{Hongyu Zhou}, \bibinfo{person}{Xin Zhou},
  \bibinfo{person}{Zhiwei Zeng}, \bibinfo{person}{Lingzi Zhang}, {and}
  \bibinfo{person}{Zhiqi Shen}.} \bibinfo{year}{2023}\natexlab{b}.
\newblock \showarticletitle{A Comprehensive Survey on Multimodal Recommender
  Systems: Taxonomy, Evaluation, and Future Directions}.
\newblock \bibinfo{journal}{\emph{arXiv preprint arXiv:2302.04473}}
  (\bibinfo{year}{2023}).
\newblock


\bibitem[Zhou et~al\mbox{.}(2023a)]%
        {zhou2023bootstrap}
\bibfield{author}{\bibinfo{person}{Xin Zhou}, \bibinfo{person}{Hongyu Zhou},
  \bibinfo{person}{Yong Liu}, \bibinfo{person}{Zhiwei Zeng},
  \bibinfo{person}{Chunyan Miao}, \bibinfo{person}{Pengwei Wang},
  \bibinfo{person}{Yuan You}, {and} \bibinfo{person}{Feijun Jiang}.}
  \bibinfo{year}{2023}\natexlab{a}.
\newblock \showarticletitle{Bootstrap latent representations for multi-modal
  recommendation}. In \bibinfo{booktitle}{\emph{Proceedings of the ACM Web
  Conference 2023}}. \bibinfo{pages}{845--854}.
\newblock


\end{thebibliography}

\appendix

\end{document}